\documentclass[11pt,onecolumn,twoside,a4paper,openany]{article}
\usepackage[dvips]{graphicx}
\usepackage{bbm}
\usepackage{amsfonts,amsmath,amssymb,graphics,psfrag}
\makeatletter
\@addtoreset{equation}{section}
\makeatother

\newcommand{\lL}{{}^{\scriptscriptstyle (\ell)}\!\!\:}

\newcommand{\lp}{\ell_p}
\newcommand{\zu}{\!\!>}
\newcommand{\auf}{\;<} 
\newcommand{\sst}{\scriptscriptstyle}
\newcommand{\st}{\scriptstyle}

\newcommand{\op}[1]{\widehat{#1}}
\newcommand{\wt}[1]{\widetilde{#1}}
\newcommand{\bet}[2]{|\beta^{#1}\,{\st #2} \zu}
\newcommand{\betc}[2]{\auf\beta^{#1}\,{\st #2}|}
\newcommand{\alp}[3]{|\alpha^{\st #1}_{#2}\,{\st #3}\zu}

\DeclareMathOperator{\sgn}{sgn}
\DeclareMathOperator{\tr}{Tr}
\def\be{\begin{equation}}
\def\ee{\end{equation}}
\def\ba{\begin{eqnarray}}
\def\ea{\end{eqnarray}}
\def\Nl{{\mathchoice
{\setbox0=\hbox{$\displaystyle\rm N$}\hbox{\hbox to0pt
{\kern0.4\wd0\vrule height0.9\ht0\hss}\box0}}
{\setbox0=\hbox{$\textstyle\rm N$}\hbox{\hbox to0pt
{\kern0.4\wd0\vrule height0.9\ht0\hss}\box0}}
{\setbox0=\hbox{$\scriptstyle\rm N$}\hbox{\hbox to0pt
{\kern0.4\wd0\vrule height0.9\ht0\hss}\box0}}
{\setbox0=\hbox{$\scriptscriptstyle\rm N$}\hbox{\hbox to0pt
{\kern0.4\wd0\vrule height0.9\ht0\hss}\box0}}}}
\def\Zl{{\mathchoice
{\setbox0=\hbox{$\displaystyle\rm Z$}\hbox{\hbox to0pt
{\kern0.4\wd0\vrule height0.9\ht0\hss}\box0}}
{\setbox0=\hbox{$\textstyle\rm Z$}\hbox{\hbox to0pt
{\kern0.4\wd0\vrule height0.9\ht0\hss}\box0}}
{\setbox0=\hbox{$\scriptstyle\rm Z$}\hbox{\hbox to0pt
{\kern0.4\wd0\vrule height0.9\ht0\hss}\box0}}
{\setbox0=\hbox{$\scriptscriptstyle\rm Z$}\hbox{\hbox to0pt
{\kern0.4\wd0\vrule height0.9\ht0\hss}\box0}}}}
\def\Ql{{\mathchoice
{\setbox0=\hbox{$\displaystyle\rm Q$}\hbox{\hbox to0pt
{\kern0.4\wd0\vrule height0.9\ht0\hss}\box0}}
{\setbox0=\hbox{$\textstyle\rm Q$}\hbox{\hbox to0pt
{\kern0.4\wd0\vrule height0.9\ht0\hss}\box0}}
{\setbox0=\hbox{$\scriptstyle\rm Q$}\hbox{\hbox to0pt
{\kern0.4\wd0\vrule height0.9\ht0\hss}\box0}}
{\setbox0=\hbox{$\scriptscriptstyle\rm Q$}\hbox{\hbox to0pt
{\kern0.4\wd0\vrule height0.9\ht0\hss}\box0}}}}
\def\Rl{{\mathchoice
{\setbox0=\hbox{$\displaystyle\rm R$}\hbox{\hbox to0pt
{\kern0.4\wd0\vrule height0.9\ht0\hss}\box0}}
{\setbox0=\hbox{$\textstyle\rm R$}\hbox{\hbox to0pt
{\kern0.4\wd0\vrule height0.9\ht0\hss}\box0}}
{\setbox0=\hbox{$\scriptstyle\rm R$}\hbox{\hbox to0pt
{\kern0.4\wd0\vrule height0.9\ht0\hss}\box0}}
{\setbox0=\hbox{$\scriptscriptstyle\rm R$}\hbox{\hbox to0pt
{\kern0.4\wd0\vrule height0.9\ht0\hss}\box0}}}}
\def\Cl{{\mathchoice
{\setbox0=\hbox{$\displaystyle\rm C$}\hbox{\hbox to0pt
{\kern0.4\wd0\vrule height0.9\ht0\hss}\box0}}
{\setbox0=\hbox{$\textstyle\rm C$}\hbox{\hbox to0pt
{\kern0.4\wd0\vrule height0.9\ht0\hss}\box0}}
{\setbox0=\hbox{$\scriptstyle\rm C$}\hbox{\hbox to0pt
{\kern0.4\wd0\vrule height0.9\ht0\hss}\box0}}
{\setbox0=\hbox{$\scriptscriptstyle\rm C$}\hbox{\hbox to0pt
{\kern0.4\wd0\vrule height0.9\ht0\hss}\box0}}}}
\def\Co{{\mathchoice
{\setbox0=\hbox{$\displaystyle\rm C$}\hbox{\hbox to0pt
{\kern0.4\wd0\vrule height0.9\ht0\hss}\box0}}
{\setbox0=\hbox{$\textstyle\rm C$}\hbox{\hbox to0pt
{\kern0.4\wd0\vrule height0.9\ht0\hss}\box0}}
{\setbox0=\hbox{$\scriptstyle\rm C$}\hbox{\hbox to0pt
{\kern0.4\wd0\vrule height0.9\ht0\hss}\box0}}
{\setbox0=\hbox{$\scriptscriptstyle\rm C$}\hbox{\hbox to0pt
{\kern0.4\wd0\vrule height0.9\ht0\hss}\box0}}}}
\def\Hl{{\mathchoice
{\setbox0=\hbox{$\displaystyle\rm H$}\hbox{\hbox to0pt
{\kern0.4\wd0\vrule height0.9\ht0\hss}\box0}}
{\setbox0=\hbox{$\textstyle\rm H$}\hbox{\hbox to0pt
{\kern0.4\wd0\vrule height0.9\ht0\hss}\box0}}
{\setbox0=\hbox{$\scriptstyle\rm H$}\hbox{\hbox to0pt
{\kern0.4\wd0\vrule height0.9\ht0\hss}\box0}}
{\setbox0=\hbox{$\scriptscriptstyle\rm H$}\hbox{\hbox to0pt
{\kern0.4\wd0\vrule height0.9\ht0\hss}\box0}}}}
\def\Ol{{\mathchoice
{\setbox0=\hbox{$\displaystyle\rm O$}\hbox{\hbox to0pt
{\kern0.4\wd0\vrule height0.9\ht0\hss}\box0}}
{\setbox0=\hbox{$\textstyle\rm O$}\hbox{\hbox to0pt
{\kern0.4\wd0\vrule height0.9\ht0\hss}\box0}}
{\setbox0=\hbox{$\scriptstyle\rm O$}\hbox{\hbox to0pt
{\kern0.4\wd0\vrule height0.9\ht0\hss}\box0}}
{\setbox0=\hbox{$\scriptscriptstyle\rm O$}\hbox{\hbox to0pt
{\kern0.4\wd0\vrule height0.9\ht0\hss}\box0}}}}
\title{Consistency Check on Volume and Triad Operator Quantisation\\
in\\ Loop Quantum Gravity I}
\author{
K. 
Giesel\thanks{Kristina.Giesel@aei.mpg.de, kgiesel@perimeterinstitute.ca}
~~ and ~
T. 
Thiemann\thanks{Thomas.Thiemann@aei.mpg.de, tthiemann@perimeterinstitute.ca}\\
\\
MPI f. Gravitationsphysik, Albert-Einstein-Institut, \\
           Am M\"uhlenberg 1, 14476 Potsdam, Germany\\
\\
and\\
\\
Perimeter Institute for Theoretical Physics, \\ 
31 Caroline Street N, Waterloo, ON N2L 2Y5, Canada}
\date{{\small Preprint AEI-2005-100}}

\begin{document}
\maketitle
\begin{abstract}
The volume operator plays a pivotal role for the quantum dynamics of Loop
Quantum Gravity (LQG). It is essential in order to construct Triad 
operators that enter the Hamiltonian constraint and which become densely
defined operators on the full Hilbert space even though in the classical 
theory the triad becomes singular when classical GR breaks down. 

The expression for the volume and triad operators derives from the 
quantisation of the fundamental electric flux operator of LQG by a 
complicated 
regularisation procedure. In fact, there are two inequivalent volume 
operators available in the literature and, moreover, both operators are
unique only up to a finite, multiplicative constant which should be viewed 
as a regularisation ambiguity.

Now on the one hand, classical 
volumes and triads can be expressed directly in terms of 
fluxes and this fact was used to construct the corresponding 
volume and triad operators. 
On the other hand, fluxes can be expressed in terms of triads and 
therefore one can also view the volume operator as fundamental and 
consider the flux operator as a derived operator. 

In this paper we mathematically implement this second point of view and 
thus can examine whether the volume, triad and flux quantisations are 
consistent with each other. The results of this consistency analysis are 
rather surprising. Among other findings we show: 1. The regularisation 
constant can 
be uniquely fixed. 2. One of the volume operators can be ruled out as 
inconsistent. 3. Factor ordering ambiguities in the definition of triad 
operators are immaterial for the classical limit of the derived flux 
operator. 

The results of this paper show that within full LQG triad operators are 
consistently quantized.\newline
In this paper we merely present ideas and results of the consistency check. In a companion paper we supply detailed proofs.
\end{abstract}
\newpage


\newpage
\section{Introduction}
\label{s1}                            %
The major unresolved problem in Loop Quantum Gravity (LQG) 
(see \cite{1} for books and \cite{1a} for reviews)
is the satisfactory implementation of the quantum dynamics which in turn 
is governed by the Wheeler -- DeWitt quantum constraint \cite{2,2a}, also
called the Hamiltonian constraint.
The volume operator \cite{3,4} plays a pivotal role for the very 
definition of the Hamiltonian constraint because with its help one can
quantize triad functions which enter the classical expression for the 
Hamiltonian constraint. In particular, one takes advantage of a Poisson bracket identity between the triads and the Poisson bracket among the Ashtekar connection and the classical scalar volume function. 
Namely, one uses the methods of canonical 
quantisation, that is, the axioms of quantum mechanics, according to which 
Poisson brackets between  
classical functions are turned into 
commutators between the corresponding operators divided by $i\hbar$,
at least to lowest order in $\hbar$.
Quite surprisingly, the Hamiltonian constraint operator and similarly 
also length operators \cite{5} are then densely 
defined on the full kinematical Hilbert space of LQG although 
the classical triad becomes singular in physically relevant situations 
such as black holes or the big bang.

While playing such a distinguished role for the most important open 
problem of LQG, the volume operator and thus the derived triad operators 
have never been critically examined concerning their physical correctness 
and mathematical consistency. By the first we mean that it has never been 
shown within full LQG that the volume operator has the correct classical 
limit with respect to suitably chosen kinematical semiclassical 
states, for example those constructed in \cite{6}\footnote{This has been 
achieved, so far, only within a certain approximation 
\cite{7} which essentially consists in replacing $SU(2)$ by $U(1)^3$.
The necessary calculations in the Non Abelean case are 
much more complictaed due to two reasons: 1. The spectrum of the volume 
operator is not available in analytical form and 2. the semiclassical
analysis is calculationally more difficult. However, work is now in 
progress in order to fill this gap.}. 

By the second we mean the following:
The fundamental kinematical algebra $\mathfrak{A}$ on which LQG is based is
the Holonomy -- Flux algebra \cite{8} and its representation theory 
together with background independence leads to a unique kinematical 
Hilbert space \cite{9}. Now classically the volume and triad can 
be written as limits of functions of the flux. To implement them at the 
quantum level, one has to go through a complicated regularisation 
procedure and to take the limit. It is surprising that the resulting 
operators are densely defined at all and in fact have a discrete spectrum
because they are highly non -- polynomial expressions as functions of 
fluxes.
This is the payoff for background independence, since in background 
dependent formulations, such as the standard Fock representations these 
operators are too singular. On the other hand, since there is little 
experience with non Fock representations, it is not at all clear whether
the corresponding operators have anything to do with their classical 
counterpart. In fact, there are at least two ambiguities already at the 
level of the volume operator: First of all, there are in fact two 
unitarily inequivalent volume operators \cite{3,4} which come from two,
a priori equally justified background independent regularisation 
techniques. We will denote 
them by Rovelli -- Smolin (RS) and Ashtekar -- Lewandowski (AL) volume
respectively for the rest of this paper. Secondly, both volume operators 
are anyway only determined up to a multiplicative regularisation constant
$C_{reg}$ \cite{10} which remains undetermined when taking the limit, 
quite similar to finite regularisation constants that appear in 
counterterms of standard renormaisation of ordinary QFT. The ambiguity 
is further enhanced by factor ordering ambiguities once we consider 
triad operators. These ambiguities are parameterized by a spin quantum 
number $\ell=1/2,1,3/2,..$.

In this paper we will be able to remove all those ambiguities by the 
following consistency check: As we mentioned above, the volume and triad
can be considered as functions of the fluxes. But the converse is also 
true: The fluxes can be written in terms of triads and thus the volume. Is it then true that 
there exists a regularisation constant for the volume operator and a 
factor ordering of the flux operator considered as a function of the triad 
operator  or volume operator such that the corresponding alternative flux operator agrees
(at least in the correspondence limit of large eigenvalues of the volume 
operator) with the fundamental flux operator, independent of the choice 
of $\ell$? This better be possible as otherwise the inescapable 
conclusion would be that the volume operator is inconsistently 
quantised\footnote{In contrast, the triad operator follows from the 
volume 
operator by the axioms of quantum mechanics namely that Poisson brackets be replaced by commutators divided by $i\hbar$ and therefore it is not 
possible that the source of a possible problem is in the quantisation 
of the triad operator.}.

We will be able to precisely answer this question affirmatively. In more 
detail we will show:
\begin{itemize}
\item[1.] The RS volume operator is inconsistent with the flux operator, the AL volume operator 
is consistent.
\item[2.] $C_{reg}=1/48$ can be uniquely fixed, there is no other choice 
which is semiclassically acceptable. Remarkably, this is precisely the 
value that was obtained in \cite{4} by a completely different argument.
\item[3.] The choice of $\ell$ plays no role semiclassically, in fact it 
drops out of the final expression for the alternative operator altogether.
Therefore the afore mentioned factor ordering ambiguity is absent as far 
as the flux operator is concerned.
\item[4.] There is yet one more ambiguity in LQG which is already present
classically: Classically it is possible to consider the electric field 
either as a two form or as a pseudo two form. The corresponding 
sign of 
the determinant of the triad is then either encoded in the electric field 
or in the conjugate connection. One can take either point of view without 
affecting the symplectic structure of the theory. We will be able to show 
that one {\it must} consider the electric field as a pseudo two form, 
otherwise the alternative flux operator becomes the zero operator!
\item[5.] As expected, the alternative and fundamental flux operator agree 
for all values of the Immirzi parameter \cite{12}, hence it cannot be 
fixed by our analysis which is good because it has been fixed already by 
arguments coming from quantum black hole physics \cite{13}.
\item[6.] The calculations in this paper make extensive use of certain 
advances in technology concerning the matrix elements of the volume 
operator \cite{14}. Thus, our calculations provide an independent check of 
\cite{14} as well.
\item[7.] The factor ordering of the alternative flux operator is 
unique if one insists on the principle of minimality\footnote{By this we 
mean that given a function $f$ with self adjoint quantisation $\hat{f}$ 
and a multiplication operator $g$ for which $g^{-1}$
is defined everywhere on the Hilbert space, we may always consider 
instead 
$\hat{f}'=(g \hat{f} g^{-1}+\overline{g^{-1}} \hat{f} \overline{g})/2$ 
as the operator corresponding to $f$ if it has selfadjoint extensions as well. By minimalistic we mean the choice 
$g=1$. This issue is always present even in ordinary quantum mechanics
however it is usually not mentioned because one usually deals with 
polynomials and $g\not=1$ would destroy polynomiality. In GR the 
expressions are generically non -- polynomial from the outset and thus 
polynomiality is not available as a simplistic criterion. However, we may still insist on a minimal number of such kind of factor ordering ambiguities.}.  
\end{itemize}
These results show that instead of taking holonomies and fluxes as 
fundamental operators one could instead use holonomies and volumes as
fundamental operators. It also confirms that the method to quantise the 
triad developed in \cite{2} is mathematically consistent.\\
\\
This paper is organised as follows:\\
\\
In section two we review the regularisation and definition of the 
fundamental flux operator for the benefit of the reader and in order
to make the comparison with the alternative quantisation easier.\\
In section three we eplain the idea on which the construction of the usual flux operator is based on.\\
In section four we derive the classical expression for the alternative 
flux operator.\\
In section five  and six we discuss our results for the alternative flux operator and compare them with the corresponding results of the usual flux operator.\\
Finally in section seven we summarise and conclude.
\newline\newline
All the technical details and tools that are needed to perform this consistency check are providid in our companion paper \cite{GT}.
\newpage
\section{Review of the usual Flux Operator}       %
\begin{figure}[hbt]
   \center
   \psfrag{t+}{$t=+\epsilon$}
   \psfrag{t=0}{$t=0$}
   \psfrag{t-}{$t=-\epsilon$}
   \psfrag{t}{$t$}
   \psfrag{S}{$S_t$}
   \includegraphics[height=6cm]{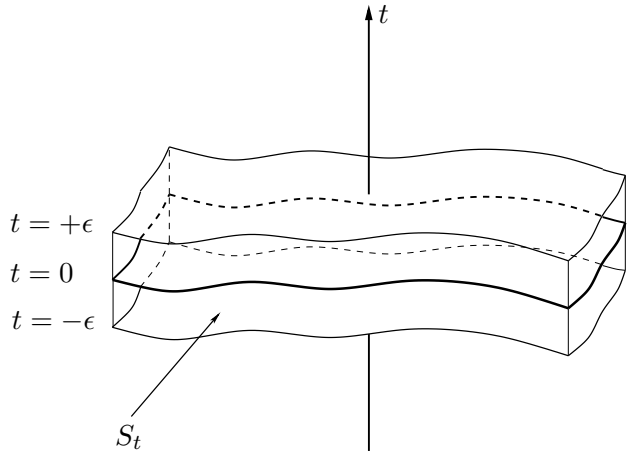}
   \caption{\label{Bild1}\small Smearing of the surface $S$ into the third dimension. We obtain an array of surfaces $S_t$ labelled by the parameter $t$ with $t\in\{-\epsilon,+\epsilon\}$. The original surface $S$ is associated with $t=0$.}
   \end{figure}

\begin{figure}[hbt]
   \center
   \psfrag{up}{up}
   \psfrag{down}{down}
   \psfrag{in}{in} 
   \psfrag{out}{out} 
   \includegraphics[height=4cm]{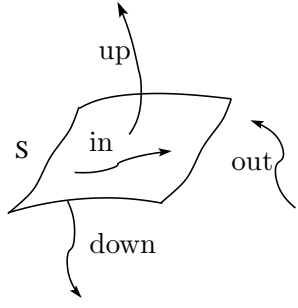}
   \caption{\label{Bild2}\small Edges of type up, down, in and out with respect to the surface $S$.}
 \end{figure}

In LQG the classical electric flux $E_k(S)$ through a surface $S$  is the integral of the densitised triad $E^a_k$ over a two surface $S$
\be
E_k(S)=\int\limits_{S}E^a_k\,n^{\st S}_a,
\ee
where $n^{\st S}_a$ is the conormal vector with respect to the surface $S$. In order to define a coresponding flux operator in the quantum theory, we have to regularise the
classical flux and then define the action of the operator on an arbitary spin network funtion (SNF) $T_{\gamma,\vec{j},\vec{m},\vec{n}}:G^{|E(\gamma)|}\to\Co$,where $G$ is the corresponding gauge group, namely $SU(2)$ in our case, as the action of the regularised expression,
denoted by $E^{\epsilon}_k(S)$ in
the limit where the regularisation  parameter $\epsilon$ is removed
\be
\label{PB}
\op{E}_k(S)T_{\gamma,\vec{m},\vec{n}}:=i\hbar\lim\limits_{\epsilon\to {\st 0}}\left\{E^{\epsilon}_k(S),T_{\gamma,\vec{m},\vec{n}}\right\}.
\ee
Here the limit is to be understood in the following way: The Poisson bracket on the right hand side of eqn (\ref{PB}) is calculated by viewing  $T_{\gamma,\vec{j},\vec{m},\vec{n}}$ as functions of smooth connections. After taking $\epsilon\to 0$ one extends the result to functions of distributional connections and thus ends up with an operator defined on the Hilbert space of LQG.\\
Classically, we have
\be
\label{PB2}
\left\{E^{\epsilon}_k(S),T_{\gamma,\vec{j},\vec{m},\vec{n}}(\{h_e(A)\}_{e\in E(\gamma)})\right\}=\sum\limits_{e\in E(\gamma)}\left\{E^{a,\epsilon}_k,(h_e)_{\sst AB}\right\}\frac{\partial T_{\gamma,\vec{j},\vec{m},\vec{n}}}{\partial(h_e)_{\sst AB}}.
\ee
Here $(h_e)_{\sst AB}$ denotes the $SU(2)$-holonomy.
 The regularisation can be implemented by smearing the two surface $S$ into the third dimension, shown in figure \ref{Bild1}, so that we get an array of surfaces $S_t$. The surface associated with $t=0$ is our original surface $S$
 \be
 \label{regflux}
 E^{\epsilon}_k(S):=\frac{1}{2\epsilon}\int\limits_{-\epsilon}^{+\epsilon}\,dt\,E_k(S_t).
 \ee
   \newline
In order to derive the action of the flux operator on an arbitrary SNF, we would have to analyse the Poisson bracket among the flux and every possible SNF. Fortunately,   
 it turnes out that each edge belonging to the associated graph $\gamma$ of $T_{\gamma,\vec{j},\vec{m},\vec{n}}$ can  be classified as (i) up, (ii) down, (iii) in and 
(iv) out with respect to the surface $S$.(See figure \ref{Bild2} for a graphical illustration.) 
An arbitrary $T_{\gamma,\vec{j},\vec{m},\vec{n}}$ contains then a particular amount of edeges of each type. Accordingly, if we have the knowledge of the Poisson bracket among the flux and any of these types 
of edges, we will be able to derive the Poisson bracket among 
$E^{\epsilon}_k$ and any arbitrary $T_{\gamma,\vec{j},\vec{m},\vec{n}}$. The calculation of the 
regularised Poisson bracket can be found e.g. in the 
second reference of \cite{1}. 
After having removed the regulator we end up with the following action of the flux operator on a SNF $T_{\gamma,\vec{j},\vec{m},\vec{n}}$
\be
\label{DerivOP}
\op{E}_k(S)T_{\gamma,\vec{m},\vec{n}}=\frac{i}{2}\,\lp^2\sum\limits_{e\in E(\gamma)}\epsilon(e,S)\left[\frac{\tau_k}{2}\right]_{AB}\frac{\partial T_{\gamma,\vec{j},\vec{m},\vec{n}}(h_{e'})_{e'\in E(\gamma)}}{\partial (h_e)_{\sst AB}},
\ee
where $\tau_k$ is related to the Pauli matrices by $\tau_k:=-i\sigma_k$. The sum is taken over all edges of the  graph $\gamma$ associated with $T_{\gamma,\vec{j},\vec{m},\vec{n}}$. The function $\epsilon(e,S)$ can take the values $\{-1,0,+1\}$ depending on the type of edge that is considered. It is +1 for edges of type up, -1 one for down and 0 for edges of type in or out. 
\newline
By introducing right invariant vector fields $X^e_k$, defined by $(X^e_k f)(h):=\frac{d}{dt}f(e^{t\tau_k}h)\Big|_{t=0}$, we can rewrite the action of the flux operator as
\be
\op{E}_k(S)T_{\gamma,\vec{j},\vec{m},\vec{n}}=\frac{i}{4}\,\lp^2\sum\limits_{e\in E(\gamma)}\epsilon(e,S)X^e_k\,T_{\gamma,\vec{j},\vec{m},\vec{n}}.
\ee
Note, the right invariant vector fields fulfill the following commutator 
relations $[X^r_e,X^s_e]=-2\epsilon_{rst}X^t_e$.
By means of introducing the self-adjoint right invariant vector field $Y^k_e:=-\frac{i}{2}X^k_e$, we achieve commutator relations for $Y^k_e$ that are similar to the one of the angular momentum operators in quantum mechanics $[Y^r_e,Y^s_e]=i\epsilon_{rst}Y^t_e$.
Therefore, we also can describe the action of $\op{E}_k(S)$ by the action of the self-adjoint right invariant vector field $Y^k_e$ on $T_{\gamma,\vec{j},\vec{m},\vec{n}}$ 
\be
\label{flY}
\op{E}_k(S)T_{\gamma,\vec{j},\vec{m},\vec{n}}=-\frac{1}{2}\lp^2\sum\limits_{e\in E(\gamma)}\epsilon(e,S)Y^k_e\,T_{\gamma,\vec{j},\vec{m},\vec{n}}.
\ee
\section{Idea of the Alternative Flux Operator} %
\label{Idea}                                    %
Our starting point will be the Poisson bracket of the Ashtekar-connection $A^j_a$ and the densitised triad $E^b_k$   given by 
\be
\left\{A^j_a(x),E^b_k(y)\right\}=\delta^3(x,y)\delta^a_b\delta^k_j
\ee
which we take as fundamental. In order to go from the ADM-formalism to the formulation in terms of Ashtekar varibales, one uses a canonical transformation. There exist two possibilities of choosing such a canonical transformation that both lead to the Poisson bracket above. These two possiblities are
\be
\begin{array}{llcl}
\mbox{I} & A^j_a=\Gamma^j_a +\gamma\sgn(\det(e))K^j_a &, &E^a_j=\frac{1}{2}\epsilon_{krs}\epsilon^{abc}e^r_b\,e^s_c\\
\mbox{II} & A^j_a=\Gamma^j_a + \gamma K^j_a &, & E^a_j=\frac{1}{2}\epsilon_{krs}\epsilon^{abc}e^r_b\,e^s_c\sgn(\det(e))
\end{array}
\ee
Here, $\Gamma^j_a$ is the SU(2)-spin connection, $K_{ab}=K^j_ae^j_b$ the extrinsic curvature (when the Gauss constraint holds) and $\gamma$ the Imirzi-parameter.
Recall again the definition of the regularised classical flux $E^{\epsilon}_k(S)$ in eqn (\ref{regflux}). Now the idea of defining an alternative regularised flux 
\be 
\label{altdtriad}
\wt{E}^{\epsilon}_k(S):=\frac{1}{2\epsilon}\int\limits_{-\epsilon}^{+\epsilon}\,dt\,\wt{E}_k(S_t)\, ;\quad\quad \wt{E}_k(S_t)=\int\limits_{S_t}E^a_k\,n^{\st S_t}_a
\ee
is to express the densitised triad $E^a_k$ in terms of the triads as above. Due to the two possibile canonical transformations, we have also two possibilities in defining an alternative densitised triad 
\be
\label{Eakdef}
E^a_k=\left\{\begin{array}{lcl} \det(e)e^a_k &=& \frac{1}{2}\epsilon_{krs}\epsilon^{abc}e^r_b\,e^s_c\\
                                         \sqrt{\det(q)}e^a_k &=& \frac{1}{2}\epsilon_{krs}\epsilon^{abc}\underbrace{\sgn(\det(e))}_{\displaystyle =:{\cal S}}e^r_b\,e^s_c\end{array}\right\}=:\left\{\begin{array}{c} E^{a,{\sst I}}_k \\ E^{a,{\sst II}}_k\end{array}\right\},
\ee
where $e^j_a$ is the cotriad related to the intrinsic metric as $q_{ab}=e^j_ae^j_b$. From now on we will use $E^{a,{\sst I}}_k$ and $E^{a,{\sst II}}_k$, respectively for the two cases.\\
So, instead of quantising the densitised triad directly, we could use the above classical identities, quantize them via the Poisson bracket identity in eqn (\ref{Pid}) and check whether both quantisation procedures are consistent. 
\newline
The main difference between these two definitions is basically a signum factor which we will denote by ${\cal S}$. From the mathematical point of view both definitions in eqn
(\ref{altdtriad}) are equally viable, thus we will keep both possibilities and emphasise the differences that occur when we choose one or the other definition. Notice however that $\det(E^{\sst I})=\det(e)^2\ge 0$ gives an anholonomic constraint which appears to be inconsistent with the definition of $\op{E}_k(S)$ as a derivative operator as in eqn (\ref{DerivOP}), as for instance pointed out in \cite{11}. We will see that this is indeed the case.\newline
Inserting the alternative expression for $E^a_k$ into eqn (\ref{regflux}), we obtain
\be
\label{altEk}
\wt{E}_k(S_t)=\left\{\begin{array}{lcrl}\int\limits_{S_t}\frac{1}{2}\epsilon_{krs}\epsilon^{abc}e^r_b\,e^s_c n^{S_t}_a &\quad ,E^{a,{\sst I}}_k=\det(e)e^a_k\\
\int\limits_{S_t}\frac{1}{2}\epsilon_{krs}\epsilon^{abc}e^r_b\,{\cal S}\,e^s_c n^{S_t}_a &\quad ,E^{a,{\sst II}}_k=\sqrt{\det(q)}e^a_k\end{array}\right\}.
\ee                 
\section{Construction of the Alternative Flux Operator}  %
\begin{figure}[htb]
   \center
    \psfrag{t}{$t$}
   \psfrag{n}{$\vec{n}^{\sst S_t}$}
   \psfrag{e'}{${\st \epsilon'}$}
   \psfrag{v}{${\st v(}{\sst \Box}{\st )}$}
   \psfrag{e3}{${\st e}_{\sst 3}{\st({\sst \Box})}$}
   \psfrag{e4}{${\st e}_{\sst 4}{\st ({\sst \Box})}$}
   \psfrag{PofS}{${\cal P}_t$ of $S_t$}
   \includegraphics[height=7cm]{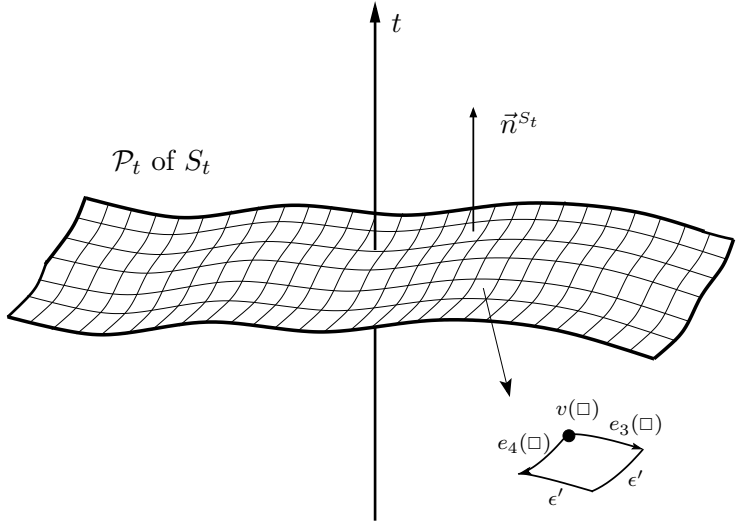}
   \caption{\label{Bild3}\small Partition ${\cal P}_t$ of the surface $S_t$ into small squares with an parameter edge length $\epsilon'$.}
    \end{figure}

The strategy for quantising the alternative Flux Operator will be as follows. As a first step we will apply the Poisson bracket identity in order to replace the triads $e^r_b$ by its associated Poisson bracket among the connection $A^r_b$ and the scalar volume funtion $V(R)=\int_R d^3x\sqrt{\det(q)}=\int_R d^3x\sqrt{|\det(E^{\sst I})|}=\int_R d^3x\sqrt{|\det(E^{\sst II})|}$. The Poisson bracket identity for the two cases is shown below
\be
\label{Pid}
\left\{A^s_b,V(R)\right\}=\left\{\begin{array}{lcl}-\frac{\kappa}{2}{\cal S}e^s_b&,& \quad E^{a,{\sst I}}_k=\frac{1}{2}\epsilon_{\sst kst}\epsilon^{\sst abc}e^s_be^t_c\\
                                                    -\frac{\kappa}{2}e^s_b&,& \quad E^{a,\sst II}_k=\frac{1}{2}\epsilon_{\sst kst}\epsilon^{\sst abc}{\cal S}e^s_be^t_c
\end{array}\right\}.
\ee 
This relation is different for the two cases, because in deriving this relation we have to use the definition of the densitised triad $E^a_k$ in terms of the triads $e^s_b$ which is different  for case I and case II. The difference between $E^{a,{\sst I}}_k$ and $E^{a,{\sst II}}_k$ is again a signum factor ${\cal S}$. Going back to eqn (\ref{altEk}) we note that there is no ${\cal S}$ in $E^{a,{\sst I}}_k$. Since we have to replace two triads $e^r_b,e^s_c$ by Poisson brackets, we get two factors of ${\cal S}$ in the case of $E^{a,{\sst I}}_k$. As  ${\cal S}^2$ is always one classically (since $q_{ab}$ is non-degenerate), it drops out in this case. In contrast, for $E^{a,{\sst II}}_k$ we have a signum factor ${\cal S}$ occuring in the alternative flux  in eqn (\ref{altEk}), but no ${\cal S}$ in the Poisson bracket identity. Accordingly, we get only one ${\cal S}$ here, that does not dissapear classically, because it can take the (constant) values $\pm 1$. Thus
\ba
\wt{E}^{\sst I}_k(S_t)&=&\frac{2}{\kappa^2}\int\limits_{S_t}\epsilon_{krs}\epsilon^{abc}\left\{A^r_b,V(R)\right\}\left\{A^s_c,V(R)\right\} n^{S_t}_a\nonumber\\
\wt{E}^{\sst II}_k(S_t)&=&\frac{2}{\kappa^2}\int\limits_{S_t}\epsilon_{krs}\epsilon^{abc}\left\{A^r_b,V(R)\right\}{\cal S}\left\{A^s_c,V(R)\right\} n^{S_t}_a
\ea
When later on we replace the classical expression by their corresponding operators, the main difference between $\wt{E}^{\sst I}_k(S_t)$ and $\wt{E}^{\sst II}_k(S_t)$ will be a so called signum operator $\op{\cal S}$. Before, we have to replace the connections $A^r_b$ by holonomies since $A^r_b$ cannot be promoted to well defined operators in the Ashtekar-Lewandowski Hilbert space ${\cal H}_{\sst AL}$. Hence, we choose a partition ${\cal P}_t$ of each surface $S_t$ into small squares of area $\epsilon'^2$. In the limit where $\epsilon'$ is small enough we are allowed to replace the connection $A^r_b$ along the edge $e_{\sst I}$ by its associated holonomy $h(e_{\sst I})$. The partition is shown in figure (\ref{Bild3}).
 Usually this is done for holonomies in the fundamental representation of $1/2$. But,  as we want to keep our construction of the alternative flux operator as general as possible and to study the effect of factor ordering ambiguities  we will consider holonomies with arbitrary representation weights $\ell$. The corresponding relation between the connection integrated along the edge $e_{\sst I}$, denoted by $A^r_{\sst I}$ from now on, and the associated holonomy is given by
 \be\label{Ah}
\left\{A^r_{\st I}({\sst {\Box}}),V(R_{v({\sst \Box})})\right\}\frac{1}{2}\pi_{\ell}(\tau_r)+o(\epsilon'^2)=+\pi_{\ell}(h_{e_{\sst I}})\left\{\pi_{\ell}(h^{-1}_{e_{\sst I}}),V(R_{v({\sst \Box})})\right\},
\ee             
whereby we indicate a representation with weight $\ell$ by $\pi_{\ell}$. 
\newline
Considering eqn (\ref{Ah}), we end up with the following classical identity for $\lL\wt{E}^{\sst I}_k(S_t)$
\ba
\label{clidI}
\lL\wt{E}^{\sst I}_k(S_t)&=&\lim_{{\cal P}_t\to S_t}\sum\limits_{\sst {\Box}\in{\cal P}_t}\epsilon_{krs}\frac{4}{\kappa^2}\left\{A^r_{\sst 3}({\sst {\Box}}),V(R_{v({\sst \Box})})\right\}\left\{A^s_{\sst 4}({\sst {\Box}}),V(R_{v({\sst \Box})})\right\}\nonumber\\    
&=&\lim_{{\cal P}_t\to S_t}\sum\limits_{\sst {\Box}\in{\cal P}_t}\frac{16}{\kappa^2}\frac{1}{\frac{4}{3}\ell(\ell+1)(2\ell+1)}\nonumber \\
&&\quad\quad\quad\quad
\tr\left(\pi_{\ell}(h_{e_{\sst 3}({\sst {\Box}})})\left\{\pi_{\ell}(h^{-1}_{e_{\sst 3}({\sst {\Box}})}),V(R_{v({\sst \Box})})\right\}\pi_{\ell}(\tau_k)\pi_{\ell}(h_{e_{\sst 4}({\sst {\Box}})})\left\{\pi_{\ell}(h^{-1}_{e_{\sst 4}({\sst {\Box}})}),V(R_{v({\sst \Box})})\right\}\right)
\ea
and in the case of $\lL\wt{E}^{\sst II}_k(S_t)$ with 
\ba
\label{clidII}
\lL\wt{E}^{\sst II}_k(S_t)&=&\lim_{{\cal P}_t\to S_t}\sum\limits_{\sst {\Box}\in{\cal P}_t}\epsilon_{krs}\frac{4}{\kappa^2}\left\{A^r_{\sst 3}({\sst {\Box}}),V(R_{v({\sst \Box})})\right\}{\cal S}\left\{A^s_{\sst 4}({\sst {\Box}}),V(R_{v({\sst \Box})})\right\}\nonumber\\
&=&\lim_{{\cal P}_t\to S_t}\sum\limits_{\sst {\Box}\in{\cal P}_t}\frac{16}{\kappa^2}\frac{1}{\frac{4}{3}\ell(\ell+1)(2\ell+1)}\nonumber \\
&&\quad\quad\quad\quad
\tr\left(\pi_{\ell}(h_{e_{\sst 3}({\sst {\Box}})})\left\{\pi_{\ell}(h^{-1}_{e_{\sst 3}({\sst {\Box}})}),V(R_{v({\sst \Box})})\right\}\pi_{\ell}(\tau_k){\cal S}\pi_{\ell}(h_{e_{\sst 4}({\sst {\Box}})})\left\{\pi_{\ell}(h^{-1}_{e_{\sst 4}({\sst {\Box}})}),V(R_{v({\sst \Box})})\right\}\right).
\ea
Here we used $\tr\left(\pi_{\ell}(\tau_r\pi_{\ell}(\tau_k))\pi_{\ell}(\tau_s)\right)=-\frac{4}{3}\ell(\ell+1)(2\ell+1)$ that is derived in appendix A of \cite{GT}.
\newline
Now, in LQG there exist two volume operators, one introduced by Rovelli and Smolin in 1994 ($\op{V}_{\sst RS}$) \cite{3} and another one published in 1995 by Ashtekar and Lewandowski ($\op{V}_{\sst AL})$ \cite{4}. Hence, we have actually in each case two different possibilities for the Poisson bracket, because we could either use $\op{V}_{\sst RS}$ or $\op{V}_{\sst AL}$.
Thus, case I as well as case II splits into two different versions of alternative fluxes
\ba
\lL\wt{E}^{\sst I}_k(S_t)&\rightarrow&\lL\wt{E}^{\sst I,AL}_k(S_t),\lL\wt{E}^{\sst I,RS}_k(S_t)\nonumber\\
\lL\wt{E}^{\sst II}_k(S_t)&\rightarrow&\lL\wt{E}^{\sst II,AL}_k(S_t),\lL\wt{E}^{\sst II,RS}_k(S_t)
\ea
From now on we will use the notation above for the four different cases. Before we apply canonical quantisation, we want to discuss the two volume operators and their differences a bit more in detail.
\subsection{The two Volume Operators in LQG}    %
 \subsubsection{The Volume Operator $\op{V}_{\sst RS}$ of Rovelli and Smolin}    %
 The idea that the volume operator acts only on vertices of a given graph was first mentioned in \cite{15}. The  first version of a volume operator can be found in \cite{3} and is given by
 \ba
\label{Voldef}
\op{V}(R)_{\gamma}&=&\int\limits_{R}\,d^3p\op{V}(p)_{\gamma}\nonumber\\
\op{V}(p)_{\gamma}&=&\lp^3\sum\limits_{v\in V(\gamma)}\delta^{(3)}(p,v)\op{V}_{v,\gamma}\nonumber\\
\op{V}^{\sst RS}_{v,\gamma}&=&\sum\limits_{I,J,K}\sqrt{
\Big| \frac{i}{8} C_{reg}
\epsilon_{ijk}X^i_{e_{\sst I}}X^j_{e_{\sst J}}X^k_{e_{\sst K}}\Big|}.
\ea
Here we sum over all triples of edges  at the vertex $v\in V(\gamma)$ of a given graph $\gamma$. $\op{V}_{\sst RS}$ is not sensitive to the orientation of the edges, thus also linearly dependent triples have to be considered in the sum. Moreover, we introduced a constant $C_{reg}\in\Rl^+$ that we will keep arbitrary for the moment and that is basically fixed by the particular regularisation scheme one chooses. 
When working with the volume operator we want to select physically relevant gauge invariant states properly. Hence, it is convenient to express our abstract angular momentum states in terms of the recoupling basis. The following identity \cite{10} holds
\be
\label{qijkDef}
\frac{1}{8}\epsilon_{ijk}X^i_{e_{\sst I}}X^j_{e_{\sst J}}X^k_{e_{\sst K}}=\frac{1}{4}[Y^2_{\sst IJ},Y^2_{\sst JK}]=:\frac{1}{4}q_{\sst IJK}^{\sst Y},
\ee
where $Y_{\sst IJ}:=Y^k_{e_{\sst I}}+Y^k_{e_{\sst J}}$ and $Y^k_{e_{\sst I}}$ denotes the self-adjoint vector field $Y^k_{e_{\sst I}}:=-\frac{i}{2}X^k_{e_{\sst I}}$. 
\newline
Consequently, we get 
\ba
\label{RSVqijk}
\op{V}(R)_{\gamma}^{\st Y,{\sst RS}}|\,J\,M\,;\,M'\zu&=&\lp^3\sum\limits_{v\in 
V(\gamma)\cap R}\sum\limits_{I<J<K}
\underbrace{\sqrt{\Big|\frac{3!i}{4} C_{reg}\op{q}_{\sst IJK}^{\sst Y}\Big|}}_{\displaystyle \op{V}^{\sst RS}_{v,\gamma}}\;\;|\,J\,M\,;\,M'\zu.
\ea
 The additional factor of $3!$ is due to the fact that we sum only over 
ordered triples $I<J<K$ now.  
The way  to calculate eigenstates and eigenvalues of $\op{V}$ is as follows. 
Let us introduce the operator $\op{Q}^{\sst Y,RS}_{v,{\sst IJK}}$ as
\be 
\label{RSQ}
\op{Q}^{\sst Y,RS}_{v,{\sst IJK}}:=\lp^6\frac{3!i}{4}C_{reg}\op{q}_{\sst IJK}^{\sst Y}
\ee
As a first step we have to calculate the eigenvalues and corresponding eigenstates for $\op{Q}^{\sst Y,RS}_{v,{\sst IJK}}$. If for example $|\phi\zu$ is an eigenstate of $\op{Q}^{\sst Y,RS}_{v,{\sst IJK}}$ with corresponding eigenvalue $\lambda$, then we obtain $\op{V}|\phi\zu=\sqrt{|\lambda|}|\phi>$. Consequently, we see that while $\op{Q}^{\sst Y,RS}_{v,{\sst IJK}}$ can have positive and negative eigenvalues, $\op{V}$ has only positive ones. Furthermore, if we consider the eigenvalues $\pm\lambda$ of $\op{Q}^{\sst Y,RS}_{v,{\sst IJK}}$ and the corresponding eigenstate $|\phi_{+\lambda}\zu,|\phi_{-\lambda}\zu$, we notice that these eigenvalues will be degenerate
  in the case of the operator $\op{V}$, as 
 $\sqrt{|+\lambda|}=\sqrt{|-\lambda|}$.
 \subsubsection{The Volume Operator $\op{V}_{\sst AL}$ of Ashtekar and Lewandowski}  %
  Another version of the volume operator which differs by the chosen regularisation scheme was defined in \cite{4}
  \ba
\label{Vqijk}
\op{V}(R)_{\gamma}^{\st Y,{\sst AL}}|\,J\,M\,;\,M'\zu&=&\lp^3\sum\limits_{v\in 
V(\gamma)\cap R}
\underbrace{\sqrt{\Big|\frac{3!i}{4} C_{reg}\sum\limits_{I<J<K}\epsilon(e_{\sst I},e_{\sst J},e_{\sst K})\,\op{q}_{\sst IJK}^{\sst Y}\Big|}}_{\displaystyle \op{V}^{\sst AL}_{v,\gamma}}\;\;|\,J\,M\,;\,M'\zu.
\ea
The major difference between $\op{V}_{\sst AL}$ and $\op{V}_{\sst RS}$ is the factor $\epsilon(e_{\sst I},e_{\sst J},e_{\sst K})$ that is sensitive to the orientation of the tangent vectors of the edges $\{e_{\sst I},e_{\sst J},e_{\sst K}\}$. $\epsilon(e_{\sst I},e_{\sst J},e_{\sst K})$ is $+1$ for right handed, $-1$ for left handed and $0$ for linearly dependent triples of edges. In the case of $\op{V}_{\sst AL}$ it is convinient to introduce an operator $\op{Q}^{\sst Y,AL}_v$ that is defined as the expression that appears inside the absolute value under the square root in $\op{V}^{\sst AL}_{v,\gamma}$
\be 
\label{Qdef}
\op{Q}^{\sst Y,AL}_v:=\lp^6\frac{3!i}{4}C_{reg}\sum\limits_{I<J<K}\epsilon(e_{\sst I},e_{\sst J},e_{\sst K})\,\op{q}_{\sst IJK}^{\sst Y}
\ee
\newline
By comparing eqn (\ref{RSVqijk}) with (\ref{Vqijk}) we notice that another difference between $\op{V}_{\sst RS}$ and $\op{V}_{\sst AL}$ is the fact that for the first one, we have to sum over the triples of edges outside the square root, while for the latter one, we sum inside the absolute value under the square root.
Apart from the difference of the sign factor, the difference in the summation will play an important role later on. 
\subsection{Factor Ordering Ambiguities}   %
The discussion in this section is valid for all four different version of the alternative flux, hence we will neglect the labels I,II,RS,AL here.
\newline
The meaning of the limit when the regularisation parameter tends to zero, i.e. $\epsilon\to 0$ in combination with the limit of the partition ${\cal P}_t$ when the edge length parameter $\epsilon'\to 0$ is explaind in detail in \cite{GT} in section 4.3. It is the same limit as taken for the usual flux operator sketched between eqn (\ref{PB}) and eqn (\ref{PB2}). We have e.g. for those $S$ which intersects $\gamma$ only in an edge of type up schematically 
\ba
\label{Limit}
\wt{E}_k(S)T&=&\lim\limits_{\epsilon\to 0}\frac{1}{2\epsilon}\int\limits_{-\epsilon}^{\epsilon}dt\sum\limits_I<T_{\tilde{t}}^{\sst I}\,|\,\wt{E}_k(S_t)\,|\,T>T_t^{\sst I} \nonumber\\
&=&\frac{1}{2}\lim\limits_{\epsilon\to 0}\sum\limits_I<T^{\sst I}_{\epsilon}\,|\,\wt{E}_k(S_{\epsilon})\,|\,T>T^{\sst I}_{\epsilon},
\ea
where $T^{\sst I}_t$ are the SNW states that contribute to $\wt{E}_k(S_T)T$. $<T^{\sst I}_{\epsilon}\,|\,\wt{E}_k(S_t)\,|\,T>$ is actually independent of $\epsilon$ and $T^{\sst I}_{\epsilon}\to T^{\sst I}_{0}$ as a function of smooth connections which can then be extended to distributional ones.

\begin{figure}[hbt]
\label{Bild4}
\center
    \psfrag{n}{$\vec{n}^{\sst S_t}$}
   \psfrag{e'}{$\epsilon'$}
   \psfrag{v}{$v({\sst \Box})$}
   \psfrag{e1}{$e_{\sst 1}({\sst \Box})$}
   \psfrag{e2}{$e_{\sst 2}({\sst \Box})$}
   \psfrag{e3}{$e_{\sst 3}({\sst \Box})$}
   \psfrag{e4}{$e_{\sst 4}({\sst \Box})$}
\includegraphics[height=5cm]{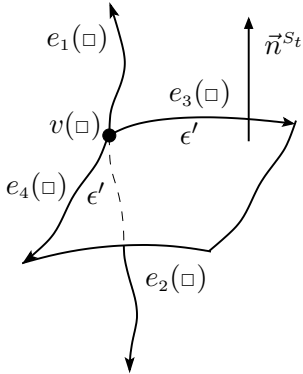}
   \caption{\small An non-vaishing contribution to the matrix element of $\lL\widehat{\wt{E}}_k$ on a given arbitrary SNF $T_{s}$, i.e. $\auf T_{s'}\,|\,\lL\widehat{\wt{E}}_k(\Box)\,|\, T_s\zu$ can only be achieved if $T_s$ contains edges of type up and/or down, respectively with respect to the surface $S_t$. Moreover, the edges $e_{\sst 3}({\sst \Box}),e_{\sst 4}({\sst \Box})$ have to be attachted to $T_s$ in this specific way.}       \end{figure}

\begin{figure}[bht]
\center
   \psfrag{n}{$\vec{n}^{\sst S_t}$}
   \psfrag{e'}{$\epsilon'$}
   \psfrag{v}{$v({\sst \Box})$}
   \psfrag{e1}{$e_{\sst 1}({\sst \Box})$}
   \psfrag{e2}{$e_{\sst 2}({\sst \Box})$}
   \psfrag{e3}{$e_{\sst 3}({\sst \Box})$}
   \psfrag{e4}{$e_{\sst 4}({\sst \Box})$}
   \psfrag{j1}{$j$}
   \psfrag{j2}{$j$}
   \psfrag{j12}{$j_{\sst 12},n_{\sst 12}$}
   \psfrag{beta}{$\bet{j_{\sst 12}}{n_{\sst 12}}$}
   \psfrag{alpha}{$\alp{J}{i}{\st M}$}
   \psfrag{J}{$J,M$}
   \psfrag{ell}{$\ell$}
   \psfrag{OP(E)}{$\lL\op{\wt{E}}_{k,{\sst tot}}(S_t)\bet{j_{\sst 12}}{n_{\sst 12}}$}      
\includegraphics[height=6cm]{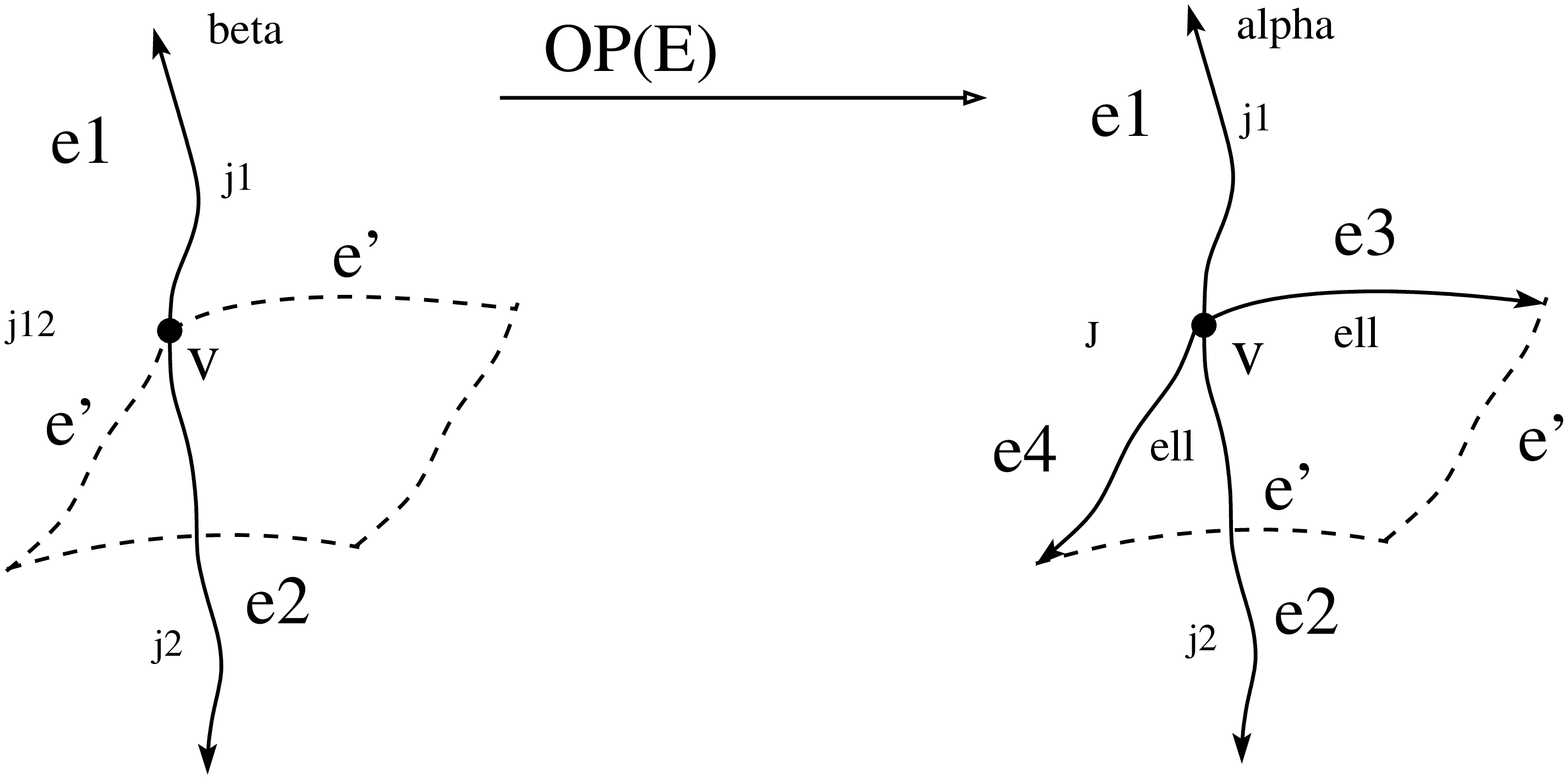}
   \caption{\label{Bild6}\small The SNF $\bet{j_{\sst 12}}{n_{\sst 12}}$ is transformed into an new SNF $\alp{J}{i}{\st M}$ by the action of $\lL\op{\wt{E}}_{k,{\sst tot}}(S_t)$.}     
   \end{figure}

It turns out that the alternative flux operator can be regularised to such an extent that each plaquette of the partition ${\cal P}_t$ of a surface $S_t$ has an intersection with only one single edge of a graph $\gamma$ associated with an arbitrary given SNF $T_{\gamma,\vec{m},\vec{n}}$. If the considered edge lies completely inside the plaquette of $S_t$ or completely outside (up to sets of dt measure zero if inside) it will lead to a trivial action of the alternative flux operator. This is in full agreement with the usual flux operator. The alternative flux operator $\lL\op{\tilde{E}}_k(S_t)$ attaches two additional edges, namely $e_{\sst 3},e_{\sst 4}$ to the edge $e$. The discussion in \cite{GT} showed that these additional edges have to lie inside the surface $S_t$ since otherwise the $\lL\op{\tilde{E}}_k(S_t)$ would have only again a trivial action, see also figure \ref{Bild4}.
 Furthermore, by attaching the additional edges $e_{\sst 3}$, $e_{\sst 4}$, $\lL\op{\tilde{E}}_k(S_t)$ creates a new vertex at a certain point of $e$ and thus divides $e$ into an edge $e_{\sst 1}$ of type up and an edge $e_{\sst 2}$ of type down with respect to $S_t$\footnote{The opposite case is also possible of course, but we will restrict our discussion to the first case and emphasize in the following disscussion where exactly the choice of $e_{\sst 1}$ as a down edge and $e_{\sst 2}$ of type up will make a difference.}. Accordinly, as for the usual flux operator, the action of the alternative flux operator is totally determined by the action on edges of type up and down with respect to the surface $S_t$.
 \newline\newline
Since we are familiar now with the action of $\lL\widehat{\wt{E}}_k(S_t)$, we are able to consider the impact of different factor orderings. Going back to the classical identity in eqn (\ref{clidI}) and eqn (\ref{clidII}) respectively and considering the results of the discussion above, we know that $\lL\widehat{\wt{E}}_k(S_t)$ adds two additional edges $e_{\sst 3},e_{\sst 4}$ to a given edge $e$ and devides this edge into an up edge $e_{\sst 1}$ and an edge of type down $e_{\sst 2}$ This is illustrated in figure \ref{Bild6}. Let us call these SNF that involves the edges $e_{\sst 1},e_{\sst 2}$ only $\bet{j_{\sst 12}}{n_{\sst 12}}$. Here we use the so called recoupling basis to express the SNF and $j_{\sst 12}$ denotes the total angular momentum to which the two edges $e_{\sst 1}$ and $e_{\sst 2}$ couple at their single vertex $v$, whereas $n_{\sst 12}$ is the associated magnetic quantum number.
 The SNF that results from $\bet{j_{\sst 12}}{n_{\sst 12}}$ by the action of $\lL\op{\wt{E}}_{k,{\sst tot}}(S_t)$ and that contains four edges $\{e_{\sst 1},e_{\sst 2},e_{\sst 3},e_ {\sst 4}\}$ will be denoted by $\alp{J}{i}{M}$. In this case $J$ is the total angular momentum, $M$ the associated magentic quantum number and $i$ as an index is needed, because with four edges more than one state exists with the same total angular momentum, but different intermediate couplings. For more details see section 5.2 and 6 in \cite{GT}.
 Now, dealing still with the classical expression in eqn (\ref{clidI}) and eqn (\ref{clidII}) respectively, we can apply the trace and rearrange the terms in a certain manner, because classically holonomies commute in order to obtain a sensible operator ordering in the quantum theory. If we consider $\lL\wt{E}^{\sst I,AL}_k(S_t)$ and $\lL\wt{E}^{\sst II,AL}_k(S_t)$ that contain $\op{V}_{\sst AL}$ we know that due to the sign factor $\epsilon(e_{\sst I},e_ {\sst J},e_{\sst K})$ in $\op{V}_{\sst AL}$ the action of $\op{V}_{\sst AL}$ on linearly dependent triples vanishes. Therefore, it has to be ensured that the two holonomies $\op{\pi}_{\ell}(h_{e_{\sst 3}})_{\sst EA}$ and $\op{\pi}_{\ell}(h_{e_{\sst 4}})_{\sst IG}$ act before $\op{V}_ {\sst AL}$ does. This restriction reduces the number of possible factor orderings down to a single one. 
 \newline
 The situation is different for $\lL\wt{E}^{\sst I,RS}_k(S_t)$ and $\lL\wt{E}^{\sst II,RS}_k(S_t)$, because here $\op{V}_{\sst RS}$ is involved, that has an non trivial action on linearly dependent triples. Consequently, more than one possible factor ordering exists. Let us discuss the number and differences of these orderings later on and restrict ourselves for $\lL\wt{E}^{\sst I,RS}_k(S_t)$ as well as $\lL\wt{E}^{\sst II,RS}_k(S_t)$ to the single ordering that is possible for $\op{V}_{AL}$, first. It turnes out that for all operators with $dt$ measure 1, the dependence of $\wt{E}_k(S_t)$ on $t$ drops out and taking the average $\frac{1}{2\epsilon}\int\limits_{-\epsilon}^{+\epsilon}dt$ becomes trivial resulting in eqn (\ref{Limit}). We will thus keep the label $S_t$  in what follows, but keep in mind that the $t$-dependence is trivial. 
\subsection{Canonical Quantisation}  %
Usually the densitised triads, appearing in the classical flux $E_k(S)$ 
 are quantised as differential operators, while holonomies are quantised 
 as multiplication operators. If we choose the alternative expression 
 $\wt{E}_k(S)$ we will instead get the scalar volume $\op{V}$ and the so 
 called signum $\op{\cal S}$ operator into our quantised expression.  The 
 properties of this $\op{\cal S}$ will be explained in more detail below. 
 Moreover, we have to replace Poisson brackets by commutators, following 
 the replacement rule $\{.\,,\,.\}\rightarrow (1/i\hbar)[.\,,\,.]$. 
 The detailed derivation of the final operator can be found in section 4 of cite{GT}.\\
 Clearly, we want the total operator to be self-adjoint, so we will calculate the adjoint of $\lL\op{\wt{E}}_k(S_t)$ and define the total and final operator to be $\lL\op{\wt{E}}_{k,{\sst tot}}(S_t)=\frac{1}{2}(\op{\wt{E}}_k(S_t)+\hat{\tilde{E^{\dagger}}}_k(S_t))$ that is self-adjoint by construction. Hence, the final operator for $\op{V}_{\sst RS}$ which we will use through the calculation of this paper is given by
 \ba
\label{RSEktot}
\lL\op{\wt{E}}^{\sst I/II,RS}_{k,{\sst tot}}(S_t)&=&\lim_{{\cal P}_t\to S_t}
\sum\limits_{\sst {\Box}\in{\cal P}_t}
\frac{8\,\lp^{-4}(-1)^{2\ell}}{\frac{4}{3}\ell(\ell+1)
(2\ell+1)}\pi_{\ell}(\tau_k)_{\sst CB}\pi_{\ell}(\epsilon)_{\sst EI}
\nonumber \\
&&\quad\quad\quad\quad
\Big\{+\pi_{\ell}(\epsilon)_{\sst FC}\left[\op{\pi}_{\ell}(h_{e_{\sst 4}})_{\sst FG}\right]^{\dagger}\left[\left[\op{\pi}_{\ell}(h_{e_{\sst 3}})_{\sst BA}\right]^{\dagger},\op{V}_{\sst RS}\right]\fbox{$\op{\cal S}$}\left[\op{V}_{\sst RS},\op{\pi}_{\ell}(h_{e_{\sst 4}})_{\sst IG}\right]\op{\pi}_{\ell}(h_{e_{\sst 3}})_{\sst EA}\nonumber\\
&&\quad\quad\quad\quad\quad
 -\pi_{\ell}(\epsilon)_{\sst FB}\left[\op{\pi}_{\ell}(h_{e_{\sst 3}})_{\sst IG}\right]^{\dagger}\left[\left[\op{\pi}_{\ell}(h_{e_{\sst 4}})_{\sst EA}\right]^{\dagger},\op{V}_{\sst RS}\right]\fbox{$\op{\cal S}$}\left[\op{V}_{\sst RS},\op{\pi}_{\ell}(h_{e_{\sst 3}})_{\sst FG}\right]\op{\pi}_{\ell}(h_{e_{\sst 4}})_{\sst CA}\Big\},
\ea
whereby we used the identity $\op{\pi}_{\ell}(h^{-1}_{e_{\sst I}})_{\sst AB}=\left[\op{\pi}_{\ell}(h_{e_{\sst I}})_{\sst BA}\right]^{\dagger}$, $\pi_{\ell}(\epsilon)_{\sst AB}=(-1)^{\ell-A}\delta_{\sst A+B=0}$ and the definition of the Planck length $\lp^{-4}:=(\hbar\kappa)^{-2}$. The box surounding the signum operator $\op{\cal S}$ should indicate that it is contained in the operator in case II, while it is not in case I.
\newline
 Considering the operator $\op{V}_{\sst AL}$, we know that for each commutator only one term will contribute, because otherwise we cannot construct linearly independet triples of edges since $\{e_{\sst 1},e_{\sst 2},e_{\sst 3/4}\}$ are linearly dependent. Therefore in the case of $\op{V}_{\sst AL}$ we obtain the following final expression 
\ba
\label{Ektot}
\lL\op{\wt{E}}^{\sst I/II,AL}_{k,{\sst tot}}(S_t)&=&\lim_{{\cal P}_t\to S_t}
\sum\limits_{\sst {\Box}\in{\cal P}_t}
\frac{8\,\lp^{-4}(-1)^{2\ell}}{\frac{4}{3}\ell(\ell+1)
(2\ell+1)}\pi_{\ell}(\tau_k)_{\sst CB}\pi_{\ell}(\epsilon)_{\sst EI}
\nonumber \\
&&\quad\quad\quad\quad
\Big\{+\pi_{\ell}(\epsilon)_{\sst FC}\left[\op{\pi}_{\ell}(h_{e_{\sst 4}})_{\sst FG}\right]^{\dagger}\left[\op{\pi}_{\ell}(h_{e_{\sst 3}})_{\sst BA}\right]^{\dagger}\op{V}_{\sst AL}\fbox{$\op{\cal S}$}\op{V}_{\sst AL}\op{\pi}_{\ell}(h_{e_{\sst 4}})_{\sst IG}\op{\pi}_{\ell}(h_{e_{\sst 3}})_{\sst EA}\nonumber\\
&&\quad\quad\quad\quad\quad
 -\pi_{\ell}(\epsilon)_{\sst FB}\left[\op{\pi}_{\ell}(h_{e_{\sst 4}})_{\sst IG}\right]^{\dagger}\left[\op{\pi}_{\ell}(h_{e_{\sst 3}})_{\sst EA}\right]^{\dagger}\op{V}_{\sst AL}\fbox{$\op{\cal S}$}\op{V}_{\sst AL}\op{\pi}_{\ell}(h_{e_{\sst 4}})_{\sst FG}\op{\pi}_{\ell}(h_{e_{\sst 3}})_{\sst CA}\Big\}.
\ea
Here again for case II the signum operator is inlcuded, whereas in case I it is not.
\newline\newline
Next, we want to calculate the matrix elements $\auf \beta^{j_{\sst 12}}\,\wt{m}_{\sst 12}\,|\lL\op{\wt{E}}_{k,{\sst tot}}(S_t)|\,\beta^{j_{\sst 12}}\,m_{\sst 12}\zu$ of all four versions $\lL\op{\tilde{E}}^{\sst I,AL}_{k,{\sst tot}}(S_t)$, $\lL\op{\tilde{E}}^{\sst I,RS}_{k,{\sst tot}}(S_t)$, $\lL\op{\tilde{E}}^{\sst II,AL}_{k,{\sst tot}}(S_t)$ and $\lL\op{\tilde{E}}^{\sst II,RS}_{k,{\sst tot}}(S_t)$ of the new flux operator.
\newline
The action of the holonomy operators $\op{\pi}_{\ell}(h_{e_{\sst I}})_{\sst AB}$ on $|\beta^{j_{\sst 12}}\,m_{\sst 12}\zu$ can be described in the framework of angular momentum recoupling theory with the powerful tool of Clebsch-Gordan-Coefficients (CGC). The correspondence between the Ashtekar-Lewandowski Hilbert space ${\cal H}_{\sst AL}$ and the abstract angular momentum Hilbert space is discussed in section 5.1 in \cite{GT}. 
Hence, the matrix element will roughly speaking have the following structure
\ba
\lefteqn{\sum\limits_{\tilde{J},\wt{M},J,M}\sum\limits_{\tilde{a}_{\sst 3},\tilde{m}_{\tilde{a}_{\sst 3}},a_{\sst 3},m_{\sst 3}}
\auf \beta^{j_{\sst 12}}\,\wt{m}_{\sst 12}\,|\lL\op{\wt{E}}_{k,{\sst tot}}(S_t)|\,\beta^{j_{\sst 12}}\,m_{\sst 12}\zu}\nonumber\\
 &\propto&C(j_{\sst 12},\ell;\tilde{a}_3\,m_{\tilde{a}_ {\sst 3}})C^*(\tilde{j}_{\sst 12},\ell;a_{\sst 3}\,m_{a_{\sst 3}})C^*(\tilde{a}_{\sst 3},\ell;\tilde{J}\,\wt{M})C(a_{\sst 3},\ell;J\,M)
\auf \alpha^{\st \tilde{J}}_{i}\,{\st \wt{M}}\,|\,\op{O}\,|\,\alpha^{\st J}_j\,{\st M}\zu
\ea
Here $C(j_1,j_2;J\,M)$ denotes the CGC that we get if we couple the angular momenta $j_1$ and $j_2$ to a resulting angular momentum $J$ with corresponding magnetic quantum number $M$ and $\op{O}$ is an operator including the operator $\op{V}_{\sst RS}$ and $\op{V}_{\sst AL}$ respectively and $\op{\cal S}$.
The details can be found in section 5.1 of \cite{GT}. Due to the symmetry properties of the alternative flux operator that are analysed in section 5.2 of \cite{GT}, the only total angular momenta that will contribute to the final matrix element are $J=0,1$. (See section 5.2 of \cite{GT} for more explanations concerning this point.) Since the physically relevant states are gauge-invariant we choose $j_{\sst 12}=0$. The behaviour of $\lL\op{\wt{E}}_{k,{\sst tot}}(S_t)$ under gauge transformations, that is discussed in section 5.3 of \cite{GT} leads to the restriction $\tilde{j}_{\sst 12}=1$. Consequently, at the end of the day we get the following form of the matrix element of $\lL\op{\wt{E}}_{k,{\sst tot}}(S_t)$
\ba
\label{act7Ek}
\lefteqn{\auf \beta^{\st 1}\,\wt{m}_{\sst 12}\,|\lL\op{\wt{E}}_{k,{\sst tot}}(S_t)|\,\beta^{\st 0}\,{\st 0}\zu}\nonumber\\
&=&-\lim_{{\cal P}_t\to S_t}\sum\limits_{\sst {\Box}\in{\cal P}_t}\frac{8\,\lp^{-4}(-1)^{3\ell}}{\frac{4}{3}\ell(\ell+1)(2\ell+1)}
\sum\limits_{\sst B,C,F=-\ell}^{+\ell}\Big\{\pi_{\ell}(\tau_k)_{\sst CB}\nonumber\\
&&\Big[+(-1)^{\st -F}\delta_{\sst F+C,0}\sqrt{2\ell+1}\delta_{\wt{m}_{\sst 12}{\st +B+F,0}}\nonumber\\
&&\hspace{2.2cm}
\auf 1\, \wt{m}_{\sst 12}\, {\st ;}\,\ell\, {\st B}\,|\,\ell\, \wt{m}_{\sst 12}{\st +B}\zu 
\auf \ell\, \wt{m}_{\sst 12}{\st +B}\, {\st ;}\,\ell\, {\st F}\,|\,\,0\,0\zu
 \nonumber\\
&&\hspace{2.2cm}
\auf \alpha^{\st 0}_{2}\,{\st M}=\wt{m}_{\sst 12}{\st +B+F}\, {\st ;}\, \wt{m}'_{\sst 1}\,\wt{m}'_{\sst 2}\,|\,\op{O}_{\sst 1}\,|\,\alpha^{\st 0}_1\,{\st M}=0\, {\st ;}\, m'_{\sst 1}\,m'_{\sst 2}\zu\nonumber\\
&&
-(-1)^{\st -F}\delta_{\sst F+B,0}
\delta_{{\sst C+F},\wt{m}_{\sst 12}}\nonumber\\
&&\hspace{2.2cm}\auf 0\, 0\, {\st ;}\ell\, {\st C}|\ell\, {\st C}\zu  
\auf \ell\, {\st C}\, {\st ;}\ell\, {\st F}|\,1\, {\st C+F}\zu\nonumber\\
&&\hspace{2.2cm}
\Big[+\frac{\sqrt{2\ell-1}}{\sqrt{3}}
\auf \alpha^{\st 1}_{2}\,{\st M}=\wt{m}_{\sst 12}\, {\st ;} \wt{m}'_{\sst 1}\,\wt{m}'_{\sst 2}\,|\,\op{O}_{\sst 2}\,|\alpha^{1}_1\,{\st M}={\st C+F}\, {\st ;}\, m'_{\sst 1}\,m'_{\sst 2}\zu\nonumber\\
&&\,\,\,\,\hspace{2.2cm}
-\frac{\sqrt{2\ell+1}}{\sqrt{3}}
\auf \alpha^{\st 1}_{3}\,{\st M}=\wt{m}_{\sst 12}\, {\st ;} \wt{m}'_{\sst 1}\,\wt{m}'_{\sst 2}\,|\,\op{O}_{\sst 2}\,|\alpha^{1}_1\,{\st M}={\st C+F}\, {\st ;}\, m'_{\sst 1}\,m'_{\sst 2}\zu\nonumber\\
&&\,\,\,\,\hspace{2.2cm}
+\frac{\sqrt{2\ell+3}}{\sqrt{3}}
\auf \alpha^{\st 1}_{4}\,{\st M}=\wt{m}_{\sst 12}\, {\st ;} \wt{m}'_{\sst 1}\,\wt{m}'_{\sst 2}\,|\,\op{O}_{\sst 2}\,|\alpha^{1}_1\,{\st M}={\st C+F}\, {\st ;}\, m'_{\sst 1}\,m'_{\sst 2}\zu\Big]\Big\},
\ea
whereby the four different cases are encoded in the operators $\op{O}_1$ and $\op{O}_2$. Explicitly, we have
\ba
\label{O1O2}
\op{O}^{\sst I,AL}_1&=&\op{V}^2_{\sst AL}\nonumber\\
\op{O}^{\sst I,RS}_1&=&\op{V}_{q_{\sst 134}}^2+\op{V}_{q_{\sst 234}}^2+\op{V}_{q_{\sst 134}}\op{V}_{q_{\sst 234}}+\op{V}_{q_{\sst 234}}\op{V}_{q_{\sst 134}}+\op{V}_{q_{\sst 134}}\op{V}_{q_{\sst 123}}+\op{V}_{q_{\sst 124}}\op{V}_{q_{\sst 134}}+\op{V}_{q_{\sst 234}}\op{V}_{q_{\sst 123}}+\op{V}_{q_{\sst 124}}\op{V}_{q_{\sst 234}}+\op{V}_{q_{\sst 124}}\op{V}_{q_{\sst 123}}\nonumber\\
\op{O}^{\sst I,AL}_2&=&\op{V}^2_{\sst AL}\nonumber\\
\op{O}^{\sst I,RS}_2&=&\op{V}_{q_{\sst 134}}^2+\op{V}_{q_{\sst 234}}^2+\op{V}_{q_{\sst 234}}\op{V}_{q_{\sst 134}}+\op{V}_{q_{\sst 134}}\op{V}_{q_{\sst 234}}+\op{V}_{q_{\sst 123}}\op{V}_{q_{\sst 134}}+\op{V}_{q_{\sst 134}}\op{V}_{q_{\sst 124}}+\op{V}_{q_{\sst 123}}\op{V}_{q_{\sst 234}}+\op{V}_{q_{\sst 234}}\op{V}_{q_{\sst 124}}+\op{V}_{q_{\sst 123}}\op{V}_{q_{\sst 124}}\nonumber\\
\op{O}^{\sst II,AL}_1&=&\op{V}_{\sst AL}\op{\cal S}\op{V}_{\sst AL}\nonumber\\
\op{O}^{\sst II,RS}_1&=&+\op{V}_{q_{\sst 134}}\op{\cal S}\op{V}_{q_{\sst 134}}+\op{V}_{q_{\sst 234}}\op{\cal S}\op{V}_{q_{\sst 234}}+\op{V}_{q_{\sst 134}}\op{\cal S}\op{V}_{q_{\sst 234}}+\op{V}_{q_{\sst 234}}\op{\cal S}\op{V}_{q_{\sst 134}}+\op{V}_{q_{\sst 134}}\op{\cal S}\op{V}_{q_{\sst 123}}+\op{V}_{q_{\sst 124}}\op{\cal S}\op{V}_{q_{\sst 134}}+\op{V}_{q_{\sst 234}}\op{\cal S}\op{V}_{q_{\sst 123}}\nonumber\\
&&+\op{V}_{q_{\sst 124}}\op{\cal S}\op{V}_{q_{\sst 234}}+\op{V}_{q_{\sst 124}}\op{\cal S}\op{V}_{q_{\sst 123}}\nonumber\\
\op{O}^{\sst II,AL}_2&=&\op{V}_{\sst AL}\op{\cal S}\op{V}_{\sst AL}\nonumber\\
\op{O}^{\sst II,RS}_1&=&+\op{V}_{q_{\sst 134}}\op{\cal S}\op{V}_{q_{\sst 134}}+\op{V}_{q_{\sst 234}}\op{\cal S}\op{V}_{q_{\sst 234}}+\op{V}_{q_{\sst 234}}\op{\cal S}\op{V}_{q_{\sst 134}}+\op{V}_{q_{\sst 134}}\op{\cal S}\op{V}_{q_{\sst 234}}+\op{V}_{q_{\sst 123}}\op{\cal S}\op{V}_{q_{\sst 134}}+\op{V}_{q_{\sst 134}}\op{\cal S}\op{V}_{q_{\sst 124}}+\op{V}_{q_{\sst 123}}\op{\cal S}\op{V}_{q_{\sst 234}}\nonumber\\
&&+\op{V}_{q_{\sst 234}}\op{\cal S}\op{V}_{q_{\sst 124}}+\op{V}_{q_{\sst 123}}\op{\cal S}\op{V}_{q_{\sst 124}},
\ea
whereby we used the notation $\op{V}_{q_{\sst IJK}}$ for $\op{V}_{\sst RS}=\op{V}_{q_{\sst 134}}+\op{V}_{q_{\sst 234}}+\op{V}_{q_{\sst 124}}+\op{V}_{q_{\sst 123}}$ when only the triple $\{e_{\sst I},e_{\sst J},e_{\sst K}\}$ contributes to $\op{V}_{\sst RS}$. The derivation of the various versions of  $\op{O}_1,\op{O}_2$ can be found in \cite{GT}. These 8 versions result from taking into account a) the two volume operators, b) the signum operator ${\cal S}$ or not and c) the adjoint or not in eqn (\ref{act7Ek}).
\newline
\newline
Before we will discuss our results in the next sections, we want to explain a bit more in detail what we mean by say the consistency check is affirmative or not. We managed to implement an alternative flux operator $\lL\op{\tilde{E}}_{k,{\sst tot}}(S_t)$ in four different versions $\lL\op{\tilde{E}}^{\sst I,AL}_{k,{\sst tot}}(S_t)$, $\lL\op{\tilde{E}}^{\sst I,RS}_{k,{\sst tot}}(S_t)$, $\lL\op{\tilde{E}}^{\sst II,AL}_{k,{\sst tot}}(S_t)$ and $\lL\op{\tilde{E}}^{\sst II,RS}_{k,{\sst tot}}(S_t)$. The usual flux operator is quantised as a differential operator in the standard way. The alternative flux operator $\lL\op{\tilde{E}}_{k,{\sst tot}}(S_t)$ is quantised via the Poisson bracket identity in eqn (\ref{Pid}) analogous to the quantisation of the Hamiltonian Constraint. If these two methods of quantisation are mathematically consistent with each other, the action of $\op{E}_k(S)$ and the one of $\lL\op{\tilde{E}}_{k,{\sst tot}}(S_t)$ should only differ by a constant, namely
\be
\op{E}_k(S)\bet{0}{0}=C(j,\ell)C_{reg}\lL\op{\tilde{E}}^{\st 0}_{k,{\sst tot}}(S)\bet{0}{0}=C(j,\ell)C_{reg}\sum\limits_{\wt{m}_{\sst 12}} 
\betc{1}{\wt{m}_{\sst 12}}\,\lL\op{\wt{E}}^{\st 0}_{k,{\sst tot}}(S)\,\bet{0}{0}\bet{1}{\wt{m}_{\sst 12}},
\ee
whereby $\lL\op{\tilde{E}}^{\st 0}_{k,{\sst tot}}(S)$ denotes the alternative flux operator $\lL\op{\tilde{E}}_{k,{\sst tot}}(S)$ where $C_{reg}$ has been replaced by 1.
The constant $C(j,\ell)$ might depend on the spin labels $j,\ell$ of the edges, but at least semiclassically the dependence on the spin label $j$ and $\ell$ must disappear since otherwise the behaviour of $\op{E}_k(S)$ and $\lL\op{\tilde{E}}_{k,{\sst tot}}(S)$ in the correspondence limit of large $j$ would disagree. Notice that also a dependence of $\lim\limits_{j\to\infty}C(j,\ell)=:C(\ell)$ on $\ell$ is inacceptable, because classically the flux is independent of the factor ordering ambiguity $\ell$. Moreover this constant will fix the ambiguity in the volume operator $C_{reg}$  that is due to regularisation as we will see later. Hence, $\lim\limits_{j\to\infty}C(j,\ell)=1/C_{reg}=\mbox{const}$ as will be discussed more in detail in section \ref{C2AL}.
\section{Case I: Results for $\lL\op{\tilde{E}}^{\sst I,RS}_{k,{\sst tot}}(S_t)$ and $\lL\op{\tilde{E}}^{\sst I,AL}_{k,{\sst tot}}(S_t)$}  %
\subsection{Calculations for $\lL\op{\tilde{E}}^{\sst I,AL}_{k,{\sst tot}}(S_t)$}  %
For technical reasons, we consider only a spin label $\ell=0.5,1$, because higher spin labels cannot computed analytically anymore. Fortunately, the main properties of this case already occurs when considering small $\ell$. 
The detailed calculation in section 6.2 in \cite{GT} show 
\ba
\auf \alpha^{\st 0}_{2}\,{\st M}=\wt{m}_{\sst 12}\, {\st ;} \wt{m}'_{\sst 1}\,\wt{m}'_{\sst 2}\,|\,\op{O}^{\sst I,AL}_{1}\,|\alpha^{0}_1\,{\st M}={\st C+F}\, {\st ;}\, m'_{\sst 1}\,m'_{\sst 2}\zu&=&0\nonumber\\
\auf \alpha^{\st 1}_{i}\,{\st M}=\wt{m}_{\sst 12}\, {\st ;} \wt{m}'_{\sst 1}\,\wt{m}'_{\sst 2}\,|\,\op{O}^{\sst I,AL}_{\sst 2}\,|\alpha^{1}_1\,{\st M}={\st C+F}\, {\st ;}\, m'_{\sst 1}\,m'_{\sst 2}\zu&=&0,
\ea
where $i=2,3,4$. Going back to eqn (\ref{act7Ek}), we note that the vanishing of the matrix elements above has the consequence that the whole matrix element $\auf \beta^{j_{\sst 12}}\,\wt{m}_{\sst 12}\,|\lL\op{\wt{E}}^{\sst I,AL}_{k,{\sst tot}}(S_t)|\,\beta^{j_{\sst 12}}\,m_{\sst 12}\zu$ is zero. Since the action of $\lL\op{\wt{E}}^{\sst I,AL}_{k,{\sst tot}}(S_t)$ on an arbitrary SNF can be derived from exactly this matrix element, we can conlcude that $\lL\op{\wt{E}}^{\sst I,AL}_{k,{\sst tot}}(S_t)$ is the zero operator. Accordingly, $\lL\op{\wt{E}}^{\sst I,AL}_{k,{\sst tot}}(S_t)$ is not consistent with the usual flux operator.
\subsection{Calculations for $\lL\op{\tilde{E}}^{\sst I,RS}_{k,{\sst tot}}(S_t)$} %
Also for the operator $\lL\op{\wt{E}}^{\sst I,RS}_{k,{\sst tot}}(S_t)$ the analogous calculations discussed in section 6.3 of \cite{GT} yields only trivial matrix elements
\ba
\auf \alpha^{\st 0}_{2}\,{\st M}=\wt{m}_{\sst 12}\, {\st ;} \wt{m}'_{\sst 1}\,\wt{m}'_{\sst 2}\,|\,\op{O}^{\sst I,RS}_{1}\,|\alpha^{0}_1\,{\st M}={\st C+F}\, {\st ;}\, m'_{\sst 1}\,m'_{\sst 2}\zu&=&0\nonumber\\
\auf \alpha^{\st 1}_{i}\,{\st M}=\wt{m}_{\sst 12}\, {\st ;} \wt{m}'_{\sst 1}\,\wt{m}'_{\sst 2}\,|\,\op{O}^{\sst I,RS}_{\sst 2}\,|\alpha^{1}_1\,{\st M}={\st C+F}\, {\st ;}\, m'_{\sst 1}\,m'_{\sst 2}\zu&=&0,
\ea
Consequently, we can draw the same conlcusion as for $\lL\op{\wt{E}}^{\sst I,AL}_{k,{\sst tot}}(S_t)$ and state that $\lL\op{\wt{E}}^{\sst I,RS}_{k,{\sst tot}}(S_t)$ is the zero operator and therefore inconsistent with the usual flux operator. Futhermore as either $\lL\op{\wt{E}}^{\sst I,AL}_{k,{\sst tot}}(S_t)$
 nor $\lL\op{\wt{E}}^{\sst I,RS}_{k,{\sst tot}}(S_t)$ survive the consitency check, we can rule out, at least for the cases of $\ell=0.5,1$, the choice of $E^a_k=\det(e)e^a_k$ on which these operators are based on. To rule out the choice $E^a_k(S_t)=\det(e)e^a_k$ completely, we need to investigate the matrix element for arbitrary representation weights $\ell$. 
For higher values of $\ell$ the calculation cannot be done analytically any 
more. However, the results for $\ell=0.5,1$ 
indicate that there is an abstract reason which leads to the vanishing 
of the matrix elements for {\it any} $\ell$. We were not able to find such 
an abstract argument yet. However, even if that was not the case
and there would be a range of values for $\ell$ for which not all of the 
matrix elements would vanish, it is inacceptable that the classical
theory is independent of $\ell$ while the quantum theory strongly depends on 
$\ell$ in the correspondence limit of large $j$. 
\section{Case II: Results for $\lL\op{\tilde{E}}^{\sst II,RS}_{k,{\sst tot}}(S_t)$ and $\lL\op{\tilde{E}}^{\sst II,AL}_{k,{\sst tot}}(S_t)$}
\subsection{Calculations for $\lL\op{\tilde{E}}^{\sst II,AL}_{k,{\sst tot}}(S_t)$}
\label{C2AL}
Considering the case of the operator $\lL\op{\tilde{E}}^{\sst II,AL}_{k,{\sst tot}}(S_t)$, we can read of from eqn ($\ref{O1O2}$) the expressions $\op{O}_{\sst 1}=\op{V}_{\sst AL}\op{{\cal S}}\op{V}_{\sst AL}=\op{O}_{\sst 2}$. Since the signum operator $\op{{\cal S}}$ that corresponds to the classical expression ${\cal S}:=\sgn(\det(e))$ does not exist in the literature so far, we will in detail explain how the operator $\op{{\cal S}}$ has to be understood.
\subsection{The Signum Operator $\op{\cal S}$}
We are dealing now with case II meaning that the densitised triad is given by $E^{a,{\sst II}}_k={\cal S}\det(e)e^a_k$, where ${\cal S}:=\det(e)$. Applying the determinant onto $E^{a,{\sst II}}_k$, we get 
\be
\label{DefdetE}
\det(E)=\sgn(\det(e))\det(q)\quad\mbox{with}\quad \det(q)=[\det(e)]^2\ge0.
\ee
Therefore, we obtain
\be
\sgn(\det(E))=\sgn(\det(e))={\cal S}. 
\ee
In the following we want to show that ${\cal S}=\sgn(\det(E))$ can be identified with the signum of the expression inside the absolute value under the square roots in the definition of the AL-volume. For this purpose let us first discuss this issue on the classical level and afterwards go back into the quantum theory and see how the corresponding operator $\op{{\cal S}}$ is connected with the operator $\op{Q}^{\sst AL}_v$ in eqn ({\ref{Qdef}).
\newline
In order to do this let us consider eqn (\ref{clidII}). This equation contains the classical volume
$V(R_{v({\sst \Box})})$ where $R_{v({\sst \Box})}$ denotes a region centred around the vertex $v({\sst \Box})$.
\newline
The volume of such a cube is given by
\be
\label{VBox}
V(R_{v({\sst \Box})})=\int\limits_{R_{v({\sst \Box})}}\sqrt{\det(q)}d^3x=
\int\limits_{R_{v({\sst \Box})}}\sqrt{|\det(E)|}d^3x,
\ee 
where we used $\det(q)=|\det(E)|$ from eqn (\ref{DefdetE}). Introducing a parametrisation of the cube now,  we end up with
\be
V(R_{v({\sst \Box})})=\int\limits_{[-\frac{\epsilon'}{2},+\frac{\epsilon'}{2}]^3}\left|\frac{\partial X^{\st I}(u)}{\partial u_{\st J}}\right|\sqrt{|\det(E)(u)|}d^3u
=\int\limits_{[-\frac{\epsilon'}{2},+\frac{\epsilon'}{2}]^3}\left|\det(X)\right|\sqrt{|\det(E)(u)|}d^3u.
\ee
In order to be able to carry out the integral  we choose the cube $R_{v({\sst \Box})}$ small enough and thus, the volume can be approximated by
\be
\label{ResVBox}
V(R_{v({\sst \Box})})\approx\epsilon'^3\left|\det(\frac{\partial X}{\partial u})(v)\right|\sqrt{|\det(E)(v)|}.
\ee
Using the definition of $\det(E)=\frac{1}{3!}\epsilon_{abc}\epsilon^{jkl}E^a_jE^b_kE^c_l$, we can rewrite  eqn (\ref{VBox}) as
\be
V(R_{v({\sst \Box})})=\int\limits_{\sst \Box}\sqrt{\Big|\frac{1}{3!}\epsilon_{abc}\epsilon^{jkl}E^a_jE^b_kE^c_l\Big|}d^3x
\ee
If we again choose $R_{v({\sst \Box})}$ small enough and define the square surfaces of the cube as $S^{\sst I}$, we can re-express the volume integral over the densitised triads in terms of their corresponding electric fluxes through the surfaces $S^{\sst I}$
\ba
\label{Vflux}
V(R_{v({\sst \Box})})&\approx&\sqrt{\Big|\frac{1}{3!}\epsilon_{\sst IJK}\epsilon^{jkl}E_j(S^{\sst I})E_k(S^{\sst J})E_l(S^{\sst K})\Big|}.
\ea
The flux through a particular surfaces $S^{\sst I}$ is defined as
\be
\label{VFlux}
E_j(S^{\sst I})=\int\limits_{S^{\sst I}}E^a_j n^{S^{\sst I}}_a \quad\quad n^{S^{\sst I}}_a=\frac{1}{2}\epsilon^{\sst IJK}\epsilon_{abc}X^b_{,u_{\sst J}}X^c_{,u_{\sst K}}\Big|_{n^{\sst I}=0}.
\ee
Here $n^{S^{\sst I}}_a$ denotes the conormal vector associated with the surface $S^{\sst I}$. 
Regarding eqn (\ref{Vflux}) we realise that inside the absolute value in eqn (\ref{Vflux}) appears exactly the definition of $\det(E_j(S^{\sst I}))$. Therefore we get
\be
\label{VdetES}
V(R_{v({\sst \Box})})\approx\sqrt{\Big|\det(E_j(S^{\sst I}))\Big|}.
\ee
On the other hand, by taking advantage of the fact that the surfaces $S^{\sst I}$ are small enough so that the integral can be approximated by the value at the vertex times the size of the surface itself, we obtain for $\det(E_j(S^{\sst I}))$
\ba
\label{detES}
\det(E_j(S^{\sst I}))&\approx &\det(E^a_j(v)n^{S^{\sst I}}_a(v)\epsilon'^2)\nonumber\\
&=&\det(E^a_j(v))\det(n^{S^{\sst I}}_a(v))\epsilon'^6\nonumber\\
&=&\det(E(v))\det(n^{S^{\sst I}}_a(v))\epsilon'^6.
\ea
If we consider the definition of the normal vector in eqn (\ref{VFlux}),we can show the following identity
\ba
\label{detn}
\det(n^{S^{\sst I}}_a)&=&\det(X)^3\det(X^{-1})=\frac{\det(X)^3}{\det(X)}=\det(X)^2.
\ea
Inserting eqn (\ref{detn}) back into eqn (\ref{detES}) we have
\be
\label{detES2}
\det(E_j(S^{\sst I}))\approx \det(E(v))[\det(X(v))]^2\epsilon'^6
\ee
 and can conclude that eqn (\ref{VdetES}) is consistent with the usual definition of the volume in eqn (\ref{ResVBox}). 
\newline
Since we want to identify ${\cal S}:=\sgn(\det(E))$ with the signum that appears inside the absolute value under the square root in the definition of the volume, we can read off from eqn (\ref{VdetES}), that we still have to show $\sgn(\det(E))=\sgn(\det(E_j(S^{\sst I})))$. However, this can be done by means of eqn (\ref{detES2})
\ba
\sgn(\det(E_j(S^{\sst I})))&\approx &\sgn(\det(E(v))[\det(X(v))]^2\epsilon'^6)\nonumber\\
&=&\sgn(\det(E(v)))\sgn([\det(X(v))]^2)\sgn(\epsilon'^6)\nonumber\\ 
&=&\sgn(\det(E(v))).
\ea
Consequently, we must identify ${\cal S}$ with the signum that appears inside the absolute value under the square root in the definition of the volume $V$ in the classical theory, because it was precisely the expression $\det(E_j(S_{\st I}))$ that was used in the construction of the volume operator, defined as the square root of absolute value of $\det(E)$. In the quantum theory, we introduced the operator $\op{Q}^{\sst AL}_v$ in eqn (\ref{Qdef}), which is basically the expression inside the absolute value in the definition of the volume operator. Hence, it can be seen as the squared version of the volume operator that additionally contains  information about the signum of the expression inside the absolute values. Consequently, we have the operator identitity $\op{Q}^{\sst AL}_v=\op{V}_{\sst AL}\op{\cal S}\op{V}_{\sst AL}$. 
Now we will be left with the task to calculate particular matrix elements for $\op{Q}^{\sst AL}_v$ which can be done by means of the formula derived in \cite{14}.
\subsection{Calculations for $\lL\op{\tilde{E}}^{\sst II,AL}_{k,{\sst tot}}(S_t)$} %
One big advantage that comes along with the operator identity $\op{Q}^{\sst AL}_v=\op{V}_{\sst AL}\op{\cal S}\op{V}_{\sst AL}$ is that diagonalisation of the operator $\op{Q}^{\sst AL}_v$ is no longer necessary as it was in case I for $\op{V}^2_{\sst AL}$, because the operator $\lL\op{\tilde{E}}^{\sst II,AL}_{k,{\sst tot}}(S_t)$ contains only particular matrix elements of $\op{Q}^{\sst AL}_v$ that can be exactly calculated, even for arbitrary $\ell$, by means of the tools developed in \cite{14}. The details of this calculation can be found in \cite{GT} as well as the corresponding matrix elements of the usual flux operator $\op{E}_k(S)$. If we compare the results of the usual flux operator with the one of $\lL\op{\tilde{E}}^{\sst II,AL}_{k,{\sst tot}}(S_t)$ we can judge whether $\lL\op{\tilde{E}}^{\sst II,AL}_{k,{\sst tot}}(S_t)$ leads to a result consistent with the usual flux operator. It transpires 
\be
\label{CompAL}
\lL\op{\tilde{E}}^{\sst II,AL}_{k,{\sst tot}}(S)\bet{0}{0}=3!8C_{reg}\op{E}_k(S)\bet{0}{0}
\ee
Therefore the two operators differ only by a positive integer constant. As there is still the regularisation constant $C_{reg}$ in the above equation we can now fix it by requiring 
 that both operators do exactly agree with each other. In fact there is no other choice than exact agreement because the difference would be a global constant which does not
 decrease as we take the corresponding limit of large quantum numbers $j$. 
Thus, we can remove the regularisation ambiguity of the volume operator in this way and choose $C_{reg}$ to be $C_{reg}:=\frac{1}{3!8}=\frac{1}{48}$.
  \newline
  This is exactly the 
value of $C_{reg}$ that was obtained in \cite{4} by a completely different argument. Thus the geometrical interpretation of the value we have to choose for $C_{reg}$ is perfectly provided\footnote{The factor $8=2^3$ comes from the fact that during the regularisation one integrates a product of 3 $\delta-$distributions on $\Rl$ over $\Rl^+$ only. The factor $6=3!$ is due to the fact that one should sum over ordered triples of edges only.}.
\newline
Note that the consistency check holds in the full theory and not only in the semiclassical sector.
Consequently, the operator $\lL\op{\tilde{E}}^{\sst II,AL}_{k,{\sst tot}}(S_t)$ is consistent with the usual flux operator.
\subsection{Calculations for $\lL\op{\tilde{E}}^{\sst II,RS}_{k,{\sst tot}}(S_t)$}
Now, considering the operator $\lL\op{\tilde{E}}^{\sst II,RS}_{k,{\sst tot}}(S_t)$ things look differently. Here, a  quantisation that is consistent with $\op{V}_{\sst RS}$ of the signum operator $\op{\cal S}$ cannot be found due to the simple reason that $\op{V}_{\sst RS}$ in contrast to $\op{V}_{\sst AL}$ is a sum of single square roots. Hence, there is no origin for a global sign as it was in the case of $\op{V}_{\sst AL}$. (See also section 6.6.1 in \cite{GT}.) In retrospect there is a simple argument why the only possibility $\lL\op{\tilde{E}}^{\sst I,RS}_{k,{\sst tot}}(S_t)$ (since $\lL\op{\tilde{E}}^{\sst II,RS}_{k,{\sst tot}}(S_t)$ does not exist) is ruled out without further calculation:
Namely, the lack 
 of a factor of orientation in $\op{V}_{\sst RS}$, like $\epsilon(e_{\sst 
I},e_{\sst J},e_{\sst K})$ in $\op{V}_{\sst AL}$, leads to 
the following basic disagreement with the usual flux operator: 
 Suppose we had chosen the orientation of the surface $S$ in the 
opposite way. Then the type of the edge $e$ switches between up and down 
and similarly for $e_{\sst 1},e_{\sst 2}$. Then, the result of the usual flux 
 operator would differ by a minus sign. In the case of $\op{V}_{\sst AL}$ 
 we would get this minus sign as well due to $\epsilon(e_{\sst I},e_{\sst 
J},e_{\sst K})$ contained in $Q^{\sst Al}_v=\op{V}_{\sst AL}\op{\cal S}\op{V}_{\sst AL}$, whereas a change of the orientation of 
$e_{\sst 1},e_{\sst 2}$ would not modify the result of the alternative flux 
operator if we used $\op{V}_{\sst RS}$ instead, because it is not 
sensitive to the orientation of the edges. 
 Accordingly, we should stop here and draw the conclusion that $\lL\op{\tilde{E}}^{\sst II,RS}_{k,{\sst tot}}(S_t)$ is not consistent with $\op{E}_k(S)$. 
 \newline
 One might propose to artificially use $\op{V}_{\sst RS}\op{\cal S}_{\sst AL}\op{V}_{\sst RS}$ for $\lL\op{\tilde{E}}^{\sst II,RS}_{k,{\sst tot}}(S_t)$. Note, that we attached   the label $AL$ to $\op{\cal S}$ to emphasize that its quantisation is in agreement with $\op{V}_{\sst AL}$. This is artificial for the following reason. Suppose we have a classical quantity $A:=\det(E)$ and two different functions $f_1:=\sqrt{|A|}$ and $f_2:=\sgn(A)$. If we want to quantise the functions $f_1$ and $f_2$, we do this with the help of the corresponding selfadjoint operator $\op{A}$ and obtain due to the spectral theorem  $\op{f}_1=\sqrt{|\op{A}|}$ and $\op{f}_2=\sgn(\op{A})$. The product of operators $\op{V}_{\sst RS }\op{\cal S}_{\sst AL}\op{V}_{\sst RS}$ rather corresponds to $\op{g}_1=\op{A}^{\prime}$ and $\op{g}_2=\sgn(\op{A})$, because $\op{V}_{\sst RS}$ is quantised with a different regularisation scheme than $\op{S}$ is. This would only be justified if $\sqrt{|\op{A}|}$ and $\op{A}^{\prime}$ would agree semiclassically.  However they do not: If we compare the expressions for $V_{\sst AL}$ and $V_{\sst RS}$ 
then, schematically,
they are related in the following way when restricted to a vertex: 
$\op{V}_{v,\sst AL}=|\frac{3!i}{4}C_{reg}\sum\limits_{\sst I<J<K}\epsilon(e_{\sst 
I},e_{\sst J},e_{\sst K})\op{q}_{\sst IJK}|^{1/2}$ while 
$\op{V}_{v,\sst RS}=\sum\limits_{\sst I<J<K} |\frac{3!i}{4}C_{reg}\op{q}_{\sst IJK}|^{1/2}$. It is clear that apart from the sign
$\epsilon(e_{\sst I},e_{\sst J},e_{\sst K})$ the two operators can agree at most on states where
only one of the $\op{q}_{\sst IJK}$ is non vanishing (three or four valent graphs)
simply because $\sqrt{|a+b|}\not=\sqrt{|a|}+\sqrt{|b|}$ for generic real
numbers $a,b$.
\newline
Nevertheless by analysing $\lL\op{\tilde{E}}^{\sst II,RS}_{k,{\sst tot}}(S_t)$} when the artifical operator $\op{V}_{\sst RS}\op{\cal S}_{\sst AL}\op{V}_{\sst RS}$ is involved, we obtain 
\be
\label{CompRS}
\lL\op{\tilde{E}}^{\sst II,RS}_{k,{\sst tot}}(S)\bet{0}{0}=C(j,\ell)C_{reg}\op{E}_k(S)\bet{0}{0},
\ee
whereby $C(j,\ell)\in\Rl$ is a constant depending non-trivially on the spin labels $j,\ell$ in general. One can show that $C(j,\ell)\to C(\ell)$ semiclassically, i.e. in the limit of large $j$, which is shown  in appendix E and discussed in section 6.6.2 of of \cite{GT}. Hence $\lL\op{\tilde{E}}^{\sst II,RS}_{k,{\sst tot}}(S_t)$, including the artificial operator $\op{V}_{\sst RS}\op{\cal S}_{\sst AL}\op{V}_{\sst RS}$, would be consistent with $\op{E}_k(S)$ within the semiclassical regime of the theory if we chose $C_{reg}=1/C(\ell)$ and if $C(\ell)$ would be a universal constant. Unfortunately, $C(\ell)$ has a non-trivial $\ell$ -dependence which is inacceptable because it is absent in the classical theory. Moreover, we do not see any geometrical interpretation available for the chosen value of $C_{reg}$ for any value of $\ell$ in this case. 
One could possibly get rid of the $\ell$-dependence by simply cancelling the linearly dependent triples by hand from the definition of $\op{V}_{\sst RS}$. But then the so modified $\op{V}^{\prime}_{\sst RS}$ and $\op{V}_{\sst AL}$ would  practically become identical on $3-$ and $4-$valent vertices and moreover $\op{V}^{\prime}_{\sst RS}$ now depends on the differentiable structure of $\Sigma$. See more about this in the conclusion section. 
\section{Conclusion}    %
We hope to have demonstrated in this paper that at least certain 
aspects of LQG are remarkably 
tightly defined : Certain factor ordering ambiguities turn out to be 
immaterial, some regularisation schemes can be ruled out as unphysical 
once and for all. The fact that we can exclude the RS volume from now on as far as the quantum dynamics is concerned 
should not be viewed as a criticism of \cite{3} at all: The regularisation 
performed in \cite{3} is manifestly background independent, natural and 
intuitively very reasonable. It was the first pioneering paper on 
quantisation of kinematical geometrical operators in LQG and had a deep
impact on all papers that followed it. Most of the beautiful ideas 
spelled out in \cite{3} also entered the regularisation performed in 
\cite{4} and continue to be valid. One could never have guessed that 
the regularisation performed in \cite{3} leads to an inconsistent result. 
It 
took a decade to develop the necessary technology in order to perform the 
consistency check provided in this paper. Hence, the fact that we can 
define the theory more uniquely now should be taken as a strength of the 
theory and not as a weakness of \cite{3}. Interestingly, the very first paper on the volume operator \cite{15} that we are aware of does look more like $\op{V}_{\sst AL}$ rather than $\op{V}_{\sst RS}$. We speculate that if one had completed the ideas of \cite{15} one would have ended up with $\op{V}_{\sst AL}$ rather than $\op{V}_{\sst RS}$. Of course,  one could  take the viewpoint that the consistency check performed here is unnecessary, that one can just take some definition of the volume operator and not worry about triads. However, as triads prominently enter the dynamics of LQG such a point of view would render the quantum dynamics obsolete. In other words, the dynamics  and all other operators which depend on triads such as the length operator \cite{5} or spatially difeomorphism invariant operators forces us to use $\op{V}_{\sst AL}$ rather than $\op{V}_{\sst RS}$. Finally notice that one of the motivations for choosing $\op{V}_{\sst RS}$ rather than $\op{V}_{\sst AL}$ is that $\op{V}_{\sst RS}$ does not depend on the differentiable structure of $\Sigma$. Hence one could use homeomorphism rather than diffeomorphism in order to define "diffeomorphism invariant" states as advertised in \cite{GrotRov},\cite{FairRov}. This makes the Hilbert space of such states separable. However, notice that homeomorphisms are not a symmetry of the classical theory. Furthermore, there are other possibilities to arrive at a separable Hilbert space: One can decompose the Hilbert space ${\cal H}_{Diff}$ corresponding to the strict diffeomorphisms into an uncountably infinite direct sum of separable Hilbert spaces.  In the current proposals for the quantum dynamics all of these mutually isomorphic Hilbert spaces are left invariant. If the theory are left invariant by the Dirac observables, then they would be superselected and any one of them would capture the full physics of LQG.
\newline\newline
The technical details of the check, mainly displayed in our companion paper \cite{GT}, required a
substantial computational effort. There are an order of ten crucial stages 
in the calculation where things could have gone badly wrong. Following 
the details of our analysis, one sees that all the subtle issues mentioned 
must be properly taken care of in order to get an even qualitatively 
correct result. These subtleties involve, among other things: 
\begin{itemize}
\item[1.] The meaning of the limit as we remove the regulator and to 
define the alternative flux operator has to be 
understood in the same way as for the fundamental flux operator, otherwise 
the alternative flux opertor is identical to zero. This issue is discussed in full detail in section 4.3 of \cite{GT}
\item[2.] Switching from the spin network basis to the abstract angular 
momentum basis, discussed elaborately in section 5.1 of \cite{GT}, has to take care of the precise unitary map between these 
two representations of the angular momentum algebra, otherwise the two 
operators differ drastically from each other. This unitary map is not 
mentioned in the literature because for gauge invariant operators it drops 
out of the equations. However, for the non -- gauge invariant flux it has a 
large impact.
\item[3.] Very unexpectedly, the sign of the determinant of the triad  
enters the calculation in a crucial way: Classically one would expect it 
to be neglegible, especially in an orientable manifold. However, had we 
dropped it from the quantum computation then the alternative flux operator 
would again vanish identically. It is very pleasing to see that quantisation based on a pseudo vector density is ruled out. This is for the same reason that one cannot implement the momentum operator $i\hbar\frac{d}{dx}$ on $L_2(\Rl^+,dx)$.
\item[4.] A different ordering than the one we chose would have resulted 
again in the zero operator even in case II.
\item[5.] As for the fundamental flux operator one has to first smear
the alternative flux operator into the direction transversal to the 
surface under consideration. For the fundamental flux this implies that 
edges of type ``in'' are not acted on. Without this additional smearing
the fundamental flux would be ill defined. 
For the alternative flux this 
implies furthermore that the classification of edges into the types up, 
down, in and out is meaningfull at all. Indeed, one actually can define 
the alternative flux without the additional smearing. The result is well 
defined. However, it would differ drastically from the fundamental flux 
as soon as there is a vertex of valence higher than two of the graph of the 
spin network state in question within the surface. 
The additional smearing has 
the effect that with $dt$ measure one all the vertices within the sufaces 
$S_t$ are bivalent.
\item[6.] Furthermore, the analyses in section 4.3 of \cite{GT} shows that without the additional smearing we would be missing 
the crucial factor $1/2$ in eqn (\ref{Limit}) and our $C_{reg}$ would be off the value found in 
\cite{4}.
\item[7.] Just following the tedious calculations step by step as explicitly shown in \cite{GT} and 
evaluating all the CGC's involved one sees that all the subtle signs have 
to be there, everything fits only when doing the calculation with 100\%
accuracy. The calculation is therefore a highly sensitive consistency 
check.
\item[8.] All the $\ell$ dependence disappears even at small values of j. This is especially surprising 
because the classical approximation of a connection by a holonomy along a 
given path becomes worse as we let $\ell$ grow. Of course in the limit we
take paths of infinitesimal length, however, this is done {\it after}
quantization and it could have happened that the quantization is affected
by a non trivial $\ell$ dependence  which however should disappear in the
limit of large $j$.
\end{itemize}
It transpires that the reason for getting  a zero operator without the signum operator unveils a so far not appreciated symmetry of the volume operator. It would be desirable to understand the symmetry from a more abstract perspective.
\newline
This paper along with our companion paper \cite{GT} is one of the first papers that tightens the mathematical 
structure of full LQG by using the kind of consistency argument that we 
used here. Many more such checks should be performed in the future to 
remove ambiguities of LQG and to make the theory more rigid, in particular 
those connected with the quantum dynamics.\\
\\
\\
{\large Acknowledgements}\\
\\
It is our pleasure to thank Johannes Brunnemann for countless discussions 
about the volume operator.
We also would like to thank Carlo Rovelli and, especially, Lee Smolin for illuminating discussions.\newline
K.G. thanks the Heinrich-B\"oll-Stiftung for financial support. 
This research project was supported in part by a grant from NSERC of 
Canada to the Perimeter Institute for Theoretical Physics.


\begin{thebibliography}{99}
\parskip -6pt
\bibitem{1} C. Rovelli, ``Quantum Gravity'', Cambridge University
Press, Cambridge, 2004\\
T. Thiemann, ``Modern Canonical Quantum General Relativity'', Cambridge
University Press, 2005; gr-qc/0110034
\bibitem{1a} 
C. Rovelli, ``Loop Quantum Gravity", Living Rev. Rel. {\bf 1} (1998) 1,
gr-qc/9710008\\
T. Thiemann,``Lectures on Loop Quantum Gravity'', Lecture Notes in
Physics, {\bf 631} (2003) 41 -- 135, gr-qc/0210094\\
A. Ashtekar, J. Lewandowski, ``Background Independent Quantum Gravity:
A Status Report'', Class. Quant. Grav. {\bf 21} (2004) R53;
[gr-qc/0404018]\\
L. Smolin, ``An Invitation to Loop Quantum Gravity'', hep-th/0408048
\bibitem{2} 
T. Thiemann, ``Anomaly-free Formulation of non-perturbative,
four-dimensional Lorentzian Quantum Gravity", Physics Letters {\bf B380}
(1996) 257-264, [gr-qc/9606088]\\
T. Thiemann, ``Quantum Spin Dynamics (QSD)",
Class. Quantum Grav. {\bf 15} (1998) 839-73, [gr-qc/9606089];
``II. The Kernel of the Wheeler-DeWitt
Constraint Operator",
Class. Quantum Grav. {\bf 15} (1998) 875-905, [gr-qc/9606090];
``III.
Quantum Constraint Algebra and Physical Scalar Product in Quantum General
Relativity", Class. Quantum Grav. {\bf 15} (1998) 1207-1247,
[gr-qc/9705017];
``IV. 2+1 Euclidean Quantum Gravity as a model to test 3+1
Lorentzian Quantum Gravity", Class. Quantum Grav. {\bf 15} (1998)
1249-1280, [gr-qc/9705018]; ``V. Quantum Gravity as the Natural Regulator 
of 
the Hamiltonian Constraint of Matter Quantum Field Theories",
Class. Quantum Grav. {\bf 15} (1998) 1281-1314, [gr-qc/9705019]
\bibitem{2a} 
 T. Thiemann, ``The Phoenix Project: Master Constraint
Programme for Loop Quantum Gravity'', gr-qc/0305080\\
B. Dittrich, T. Thiemann, ``Testing the Master Constraint Programme for
Loop Quantum Gravity
I. General Framework'', gr-qc/0411138;
``II. Finite Dimensional Systems'', gr-qc/0411139;
``III. SL(2,R) Models'', gr-qc/0411140;
``IV. Free Field Theories'', gr-qc/0411141;
``V. Interacting Field Theories'', gr-qc/0411142
\bibitem{3}  C. Rovelli, L. Smolin,
``Discreteness of volume and area in quantum gravity",
Nucl. Phys. {\bf B442} (1995) 593, Erratum : Nucl. Phys. {\bf B456}
(1995) 734
\bibitem{4} A. Ashtekar, J. Lewandowski, ``Quantum Theory of Geometry II :
Volume Operators", Adv. Theo. Math. Phys. {\bf 1} (1997) 388-429 
\bibitem{5}  T. Thiemann, ``A Length Operator for Canonical Quantum
Gravity'', J. Math. Phys. {\bf 39} (1998) 3372-3392; [gr-qc/9606092]
\bibitem{6} T. Thiemann, ``Quantum Spin Dynamics (QSD): VII.
Symplectic Structures and Continuum Lattice Formulations of
Gauge Field Theories", Class.Quant.Grav.18:3293-3338,2001,
[hep-th/0005232]; ``Gauge Field Theory Coherent States (GCS): I.
General Properties", Class.Quant.Grav.18:2025-2064,2001,
[hep-th/0005233]\\
T. Thiemann, O. Winkler, ``Gauge Field Theory Coherent States
(GCS): II. Peakedness Properties", Class.Quant.Grav.18:2561-2636,2001,
[hep-th/0005237]; ``III. Ehrenfest Theorems",
Class. Quantum Grav. {\bf 18} (2001) 4629-4681, [hep-th/0005234];
``IV. Infinite Tensor Product and Thermodynamic Limit",
Class. Quantum Grav. {\bf 18} (2001) 4997-5033, [hep-th/0005235]\\
H. Sahlmann, T. Thiemann, O. Winkler, ``Coherent States for
Canonical Quantum General Relativity and the Infinite Tensor Product
Extension", Nucl.Phys.B606:401-440,2001;
[gr-qc/0102038]\\
T. Thiemann, ``Complexifier Coherent States for Canonical
Quantum General Relativity", gr-qc/0206037
\bibitem{7}  H. Sahlmann, T. Thiemann, ``Towards the QFT on
Curved Spacetime Limit of QGR. 1. A General Scheme'', [gr-qc/0207030];
``2. A Concrete Implementation", [gr-qc/0207031]
\bibitem{8}  H. Sahlmann, ``When do Measures on the Space of Connections
Support the Triad Operators of Loop Quantum Gravity?'', gr-qc/0207112;
``Some Comments on the Representation Theory of the Algebra Underlying
Loop Quantum Gravity'', gr-qc/0207111\\
H. Sahlmann, T. Thiemann, ``On the Superselection Theory of
the Weyl Algebra for Diffeomorphism Invariant Quantum Gauge Theories'',
gr-qc/0302090;
``Irreducibility of the Ashtekar-Isham-Lewandowski Representation'',
gr-qc/0303074\\
A. Okolow, J. Lewandowski, ``Diffeomorphism Covariant
Representations of the Holonomy Flux Algebra'', gr-qc/0302059\\
C. Fleischhack, ``Representations of the Weyl Algebra in Quantum
Geometry'', math-ph/0407006\\
J. Lewandowski, A. Okolow, H. Sahlmann, T. Thiemann, ``Uniqueness of
Diffeomorphism Invariant States on Holonomy -- Flux
Algebras'', to appear
\bibitem{9} A. Ashtekar, C.J. Isham, ``Representations of the Holonomy
Algebras of Gravity and Non-Abelean Gauge Theories",
Class. Quantum Grav. {\bf 9} (1992) 1433, [hep-th/9202053]\\
A. Ashtekar, J. Lewandowski, ``Representation
theory of analytic Holonomy $C^\star$ algebras", in ``Knots and
Quantum Gravity", J. Baez (ed.), Oxford University Press, Oxford 1994
                                                                                
\bibitem{10}  T. Thiemann, ``Closed Formula for the Matrix Elements of the
Volume Operator in Canonical Quantum Gravity'',
J. Math. Phys. {\bf 39} (1998) 3347-3371; gr-qc/9606091
%
\bibitem{11}
R.~Loll,
``Imposing det E > 0 in discrete quantum gravity,''
Phys.\ Lett.\ B {\bf 399} (1997) 227
[arXiv:gr-qc/9703033].
%
\bibitem{12} 
G. Immirzi,
``Quantum Gravity and Regge Calculus", Nucl. Phys. Proc. Suppl. {\bf 57} 
(1997) 65; [gr-qc/9701052]\\
C. Rovelli, T. Thiemann, ``The Immirzi Parameter in Quantum General 
Relativity'', Phys. Rev. {\bf D57} (1998) 1009-14; [gr-qc/9705059]

\bibitem{13} A. Ashtekar, J. C. Baez, K. Krasnov,
``Quantum Geometry of Isolated Horizons and Black Hole Entropy",
Adv.Theor.Math.Phys.4:1-94,2001, [gr-qc/0005126]\\
K. Meissner , ``Black Hole Entropy in Loop Quantum Gravity'',
Class. Quant. Grav. {\bf 21} (2004) 5245-5252; [gr-qc/0407052]\\
M. Domagala, J. Lewandowski, ``Black 
Hole Entropy from Quantum Geometry'', 
Class.Quant.Grav. {\bf 21} (2004) 5233-5244, 2004; [gr-qc/0407051]
%
\bibitem{14} J. Brunnemann, T. Thiemann, ``Simplification of the Spectral 
Analysis
of the Volume Operator in Loop Quantum Gravity'', gr-qc/0405060
\bibitem{15}
L. Smolin,  ``Recent Developments in Non-Perturbative Quantum Gravity``, hep-th/9202022                               
%
\bibitem{CommJB}
J. Brunnemann, private communication 
\bibitem{GT}
K.~Giesel, T.~Thiemann, ``Consistency Check on Volume and Triad Operator Quantisation in Loop Quantum Gravity II", gr-qc/0507037
\bibitem{FairRov}
W.~Fairbairn and C.~Rovelli,
J.\ Math.\ Phys.\  {\bf 45} (2004) 2802
[arXiv:gr-qc/0403047].
\bibitem{GrotRov}
N.~Grot and C.~Rovelli,
J.\ Math.\ Phys.\  {\bf 37}, 3014 (1996)
[arXiv:gr-qc/9604010].
\end{thebibliography}
\end{document}